\documentclass{article}
\usepackage{graphicx} % Required for inserting images
\usepackage{xcolor}
\usepackage{ulem}
\usepackage{dialogue}
\usepackage{float}
\usepackage{amsmath}
\usepackage{dsfont}
\usepackage{geometry}
\usepackage[utf8]{inputenc}
\usepackage[T1]{fontenc}

\title{Towards Industrial Convergence : Understanding the evolution of scientific norms and practices in the field of AI}

\author{
\vspace{-.2cm}
A. Houssard$^{1}$,
\\
\\
$^1${CIS , CNRS} \\
}

\date{}

\begin{document}

\maketitle

\begin{abstract}
In the field of artificial intelligence (AI) research, there seems to be a rapprochement between academics and industrial forces. The aim of this study is to assess whether and to what extent industrial domination in the field as well as the ever more frequent switch between academia and industry resulted in the adoption of industrial norms and practices by academics. Using bibliometric information and data on scientific code, we aimed to understand academic and industrial researchers' practices, the way of choosing, investing, and succeeding across multiple and concurrent artifacts. Our results show that, although both actors write papers and code, their practices and the norms guiding them differ greatly. Nevertheless, it appears that the presence of industrials in academic studies leads to practices leaning toward the industrial side, but also to greater success in both artifacts, suggesting that if convergence is, then it is passing through those mixed teams rather than through pure academic or industrial studies.
\end{abstract}

\section{Introduction}

During an exchange of invective on Twitter (now X) between Lecun and Elon Musk, the former mentioned, in a rather explicit way the four Mertonian norms of science.

\begin{quote}
    To qualify as science a piece of research must be correct and reproducible.To be correct and reproducible, it must be described in sufficient detail in a publication.
    To be ’published’ (to receive the seal of approval) the publication must be checked for correctness by reviewers.To be reproduced the publication must be widely available to the community and sufficiently interesting.\\
    \textit{Yann Lecun (Twitter / X ; May 2024)}
\end{quote}

These norms are, according to Merton, a codification of the ethos of modern science and its actors. They exist to bind scientists to normative imperatives that ensure that the knowledge they produce has real scientific value.

Although Lecun showed a certain attachment to these norms, one cannot help but notice the industrialization of the field of AI (of which he is a symbol) and, more generally, the changes that the scientific world and the system of knowledge production have undergone in the last thirty years.

In 1994, Gibbons and colleagues proposed the concept of ''Mode 2'' to describe and prescribe a shift in knowledge production. Their work identified a transition from a classical Mertonian vision of scientific production ('Mode 1'), centered around an academic world governed by its own norms and practices, producing knowledge subsequently utilized by society at large, towards a more fluid, hybrid, context-driven, and problem-focused form of knowledge production.

This idea of a radical reorganization of knowledge production has received both significant criticism and support. Empirical research, however, has revealed that despite policy shifts and increased involvement of non-academic actors in research, the traditional 'Mode 1' of knowledge production persists. Indeed, authors such as Barrier and others \cite{Barrier2014, Brunet2012,Shinn2006} have shown that while the importance of external actors has grown considerably, there's been much effort on the part of scientific communities to maintain boundaries and assert their autonomy.

These empirical findings led scholars such as Shinn \cite{Shinn2002,Shinn2002Helix} or Grossetti \cite{grossetti2000science}, to reconsider the impact of industrial participation, and drawing from Latour’s concept of "techno-science" \cite{Latour1987-ul}, these authors proposed new theoretical frameworks for understanding the evolving relationships and inter-dependencies among actors in knowledge production.

One particularly important paradigm emerging in response to Gibbons was the "triple helix" model. This framework introduced by Etzkowitz (1998) describes a tri-polarization of scientific norms, goals, and practices, encompassing academic, industrial, and state actors negotiating the norms and practices underlying knowledge production.

Recent contributions by Moore and Frickel \cite{moore2011science, Frickel2006-pk} have further refined this perspective, introducing the concept of "asymmetrical convergence" to characterize, from a political science standpoint, current demands for more applicable research the unevenness in collaboration (which, according to the authors, tend to profit more to industrial actors). 

Today, academics in fact collaborate more with other actors but also employ diverse strategies in their collaboration ranging from occasional partnerships where industrial resources are instrumentally leveraged to conduct research \cite{Shinn2006}, to employing patents as protective measures against industrial exploitation, or reusing data derived from previous collaborations for purely academic purposes \cite{Barrier2014}.

These findings, complemented by recent work by Papatsiba \cite{Papatsiba2013} and Kotiranta \cite{Kotiranta2020} highlighting how collaborations remain largely driven by scientific and academic motives, and contributions by Noordegraaf \cite{Noordegraaf2020} and Barrier \cite{Barrier2014} highlighting the efforts of academics to maintain their autonomy despite institutional pressures, demonstrate the resilience of academic norms and practices.

However, while these observations tend to highlight the persistence of the ''Mode 1'', they might not be directly transferable to the rapidly evolving field of artificial intelligence, which might challenge the resilience of the academic community. 

While initial discoveries were made in academia, most of the recent advances are industry-driven. The private sector, due to its substantial resources (both financial and computational) and its access to large databases, attracts most of the new talents and produces the most advanced and efficient models \cite{Ahmed2023}.

Considering that both the "Helix Model" approach is based on the assumption that academia retains a certain level of relevance, that there is a certain level of interdependence between actors, a domination of the means to generate scientific discoveries inevitably creates tension. 

In fact, in addition to the growing relevance of industrial productions, a convergence of industrial practices with academic ones is indicated by a number of factors. The establishment of laboratories by "Big Tech" that emulate the academic environment \cite{moore2011science}, the increasing participation in scientific conferences \cite{Ahmed2023} \cite{Hagendorff2021} \cite{ahmed2020democratization} and the dissemination of the research findings in scientific paper format are indicative of a real rapprochement.

Furthermore, industrial actors not only engage with research but thrive in the field. They secured a central position in the citation network \cite{https://doi.org/10.48550/arxiv.2312.12881} \cite{Frank2019}, recruit many of the new researchers, and even manage to attract senior academics \cite{Jurowetzki2021-rm}. 

In essence, the industry appears to exert a dominant influence over the field of AI, both scientifically and commercially. This situation has even prompted calls for a reinforcement of AI-related academic research by the very researchers who are engaged in it\cite{Frber2023} \cite{Gelles2024}.

The preponderance of an actor's influence over the other naturally questions models that assume mutual dependency, resulting in individual adjustment, and prompts us to question the current situation of the field. Raimbault \cite{Raimbault2022} uses the concept of ''industrial framing'' to describe the situation of the field of synthetic biology in France. Though the parallel between multiple industry driven fields of research could be made, we note one important difference: the field of AI has (or at least had) an important academic anchoring, contrary to Raimbault's example.\\ 

Our question consequently is : Is the academic field of AI research, facing a situation of industrial domination, completely adopts an ''industrial framing'' of research ? Put differently : Is the academic research aligning itself with the industrial one ? 

Although some studies have  identified the increasing influence of industrial actors in this field, there is still a deficit of knowledge regarding their impact on researcher practices. 

Using bibliometric data along with scientific code information, we provide a comparative analysis of the practices of academic and industrial researchers across concurrent artifacts. We specifically question the way to choose, invest, and succeed in multiple and concurrent artifacts.

More specifically, we investigated the situation of two concurrent artifacts, the scientific paper and code repositories. The first being a scientific production which generates scientific value and credibility for its author and the second being the means for the creation of technical artifacts (software and models) which have the potential for generating economic returns. 

Our results indicate that purely academic studies still dominate the literature and differ significantly from those involving industrial authors. Although it seems that the 'boundary work' is still in play, we also find that industrial actors are mostly represented in mixed teams, which make up a large part of our sample. Moreover, these mixed teams seem to align their research questions and practices with those of the industrial actors and are more successful across both scientific and technical artifacts.

\section{Literature review}

\subsection{The field of AI : An example of techno-science}

In their work on "mode 2," Gibbons et al. \cite{Gibbons1994-qi} present a new ideal type of science. Despite its speculative and radical nature, the authors summarize and articulate a significant number of inquiries, empirical findings, and observations, effectively presenting a paradigm shift from traditional academic research to a hybrid model.

Their work both analyzes and advocates a shift from the Mertonian norms, including a transition from an uninterested to an interested science, from a collegial to an open science, and so forth. Although their approach is speculative, some elements suggest that Gibbons' \& colleagues \cite{Gibbons1994-qi} aspiration is becoming a reality.   

The rise of project-based research \cite{Felt2016} \cite{Smith2023}, public/private partnership and governance \cite{Schot2018}, the emergence of the "research-technologists" \cite{Shinn2002} and then "research entrepreneurs" figure \cite{Raimbault2023}, the importance given to addressing socially relevant issues or producing applicable results \cite{Raimbault2023} \cite{Joly2019} are so many elements giving credit to the idea of a shift in the way knowledge is produced. In fact, Gibbons' \& colleagues argue in favor of moving beyond the modern university model. For the authors and other supporters of the "Mode 2", the shift from the classic academic model toward a fluid, in terms of institution, disciplines, funding, etc... mode of knowledge production is already enacted and preferable.     

Despite the numerous criticisms of the radical nature of Gibbon's concept and the lack of evidence that would allow one to speak of a paradigm shift, it has become standard practice among sociologists to consider the multiplicity of actors, demands, theories, methods, norms, and ethos involved in the production of new knowledge and technologies.\\ 

A review of the literature on the field AI reveals a multitude of actors and objectives. States and supranational institutions aim to capitalize on the economic potential of AI technologies while simultaneously regulating and funding their development \cite{Smith2023}. Companies engage in the production of hardware, software, and the delivery of services to secure their position in the evolving market \cite{jacobides2021evolutionary}.  

Although still active, academia is now a relatively minor player in this field. Indeed, it has lost a significant proportion of its personnel to private companies, operates with limited resources, and faces challenges in developing meaningful models \cite{Ahmed2023}.    

In his 2022 paper, Raimbault \cite{Raimbault2022} mentions the concept of ''industrial framing'' to qualify a discipline with strong industrial ties that has failed to institutionalize itself as an academic discipline. We have mentioned that there are some striking similarities between the field, but we must also recognize that the field of AI both pushes some of the logic mentioned and is in some ways different.

Although the field studied is presented as particularly industrialized, cooperation still relates to strategic decisions, while the academic path still allows research to be carried out. We can also see that the order of magnitude is significantly different. While Raimbault mentions that in a highly industrialized laboratory 40\% of PhD students go to industry, Ahmed notes that in AI only about 30\% of former PhD students stay in academia. Finally, it is mentioned that collaborations for the production of academic artifacts (e.g. academic papers) are still rare.

As we will show in the next section, the field of AI appears to be driving many of the observations made by researchers in fields that have traditionally had important industrial links, and this in a discipline that emerged and was dominated by academia for many decades\cite{Cardon2018}.

Considering this level of domination, one can only question the current place of academia within the field of AI. Moore et al. \cite{moore2011science} already show how, within the context of neoliberalism, an asymmetric convergence emerges, meaning a new alignment between academic and industrial logic with differentiated benefits depending on the field of research and often favorable to the industry but dismissed the idea of a total industrial domination leading to a unified system of knowledge production which we could actually observe here.

\subsection{Academy and Industry in AI research}

As stated above, academia has lost its role as the driving force in the advancement of AI technology. Additionally, many researchers observed that academia is also declining when it comes to scientific endeavors. As Ahmed et al. \cite{Ahmed2023} note, 70\% of new doctors now directly enter the industry. This trend is corroborated by Jurowetzki et al., who also report an increasing trend in the number of individuals who switch from academic to industrial institutions \cite{Jurowetzki2021-rm}. \\

The consequence of this brain drain is an over-representation of companies at major AI scientific conferences \cite{Hagendorff2021} \cite{ahmed2020democratization} and provides the industry with a central presence within the literature. While there are variations in methodology and findings across studies, it is evident that the production of company-written articles and hybrid articles (including academic and industrial authors) is growing significantly \cite{Frber2023} \cite{https://doi.org/10.48550/arxiv.2103.06312} \cite{Frank2019}. Furthermore, these processes attract an ever-increasing amount of attention from the academic community. Recent studies by Färber et al. \cite{Frber2023} and Giziski et al. \cite{https://doi.org/10.48550/arxiv.2103.06312}, respectively, show that industrial and hybrid papers are both at the top of the citation ranking and central within the citation network. 

Moreover, industrial research led to the production of some of the most relevant models, according to Ahmed \cite{Ahmed2023}, in 2023 all 10 largest models were industrially produced. While the industry's sharing of methods and results may lead many to view their participation in the research endeavor as beneficial, some authors have highlighted concerning trends in their studies. Despite the established benefit of industrial "engagement" in maintaining or increasing research production and quality (measured by the ability to obtain citations) \cite{Perkmann2021}, the openness of these studies is subject to controversy.\\

Firstly, it seems that industrial actors are influencing the direction of research towards a limited number of themes and technologies. A number of studies have observed the simultaneous expansion of industrial studies along with a "narrowing of AI research" \cite{Klinger2020}. In their work, authors such as Klinger \cite{Klinger2020} or Frank \cite{Frank2019} demonstrate how AI research, especially when industrial or hybrid, is increasingly focusing on a limited number of tasks and methods. 

In addition, authors such as Baruffaldi \cite{Baruffaldi2020} mention the potentially instrumental nature of these interactions. In fact, maintaining close relations with the academic world allows for more fluid exchange of knowledge as to maintain relevant or even cutting-edge technology.  

Third, a large "compute divide" exists between academic and industrial actors. The accessibility of computational resources required for the development of competitive models is constrained for academic actors \cite{ahmed2020democratization} \cite{Rikap2024}. While some research teams demonstrate the ability to replicate industrial advancements with limited resources, such discrepancies inevitably create obstacles for academic research \cite{Gelles2024}. 

Finally, in addition to computing capabilities, industrial actors also have access to large, privatized databases necessary to effectively train models \cite{Ahmed2023}. \\

In summary, industrial actors have attained a dominant position in the field of AI research and development due to their substantial financial resources. While their involvement is acknowledged and attracts the attention of academics, it also raises concerns. The "narrowing of AI research" and issues related to the openness and instrumentality in the usage of the research are manifestations of broader perturbation related to the industrial domination of the field. 

However, while this research is essential for mapping the field, it does not tell us much about changes in scientific labor and the underlying norms that guide its work. Our study aims to interrogate these dynamics by comparing the investment made in the concurrent artifact: the academic paper and the software development. 

In other words, is the work of the academic researcher aligned with that of its industrial counterpart in terms of the way, time and success they find in the artifact, following different objectives, and can we therefore speak of a unified system of knowledge production (Mode 2)?

\subsection{Publishing and sharing your code}

Although crucial scientific codes and software development have historically received little attention, both from researchers in the sociology of science, science and technology studies, and from scientific institutions. Today, partly because of growing concerns about the openness of science, the topic is gaining in popularity, and many studies are now questioning the quality and openness of scientific code. 

The code, as a scientific artifact\footnote{An artifact can be defined as any object that is part of the "common body of knowledge" \cite{duPlessis2008}. This concept can be understood as referring to any tangible product of scientific research : document, tool, methods, etc. usable by other members of the community.}, has been found in many fields to have relatively low heuristic value. In fact, studies conducted across various fields \cite{Trisovic2022} have revealed that for the most part, the code shared by academics is poorly maintained and frequently non-functional.

This situation can be explained by emphasis put on the paper itself, along with the citations and the prestige it brings. Latour \& Woolgar \cite{latour2013laboratory} already identified this situation and proposed a model of scientific credibility, which essentially demonstrates that for any virtuous circle to be enacted within academia, publications must be carried out. While this is not a particularly surprising finding, further studies have shown that even in fields that focus on applicability, the publication step remains crucial. 

In their work, authors such as Hessel \cite{Hessels2019} and Brun \cite{Brun2023} illustrate that, despite variations in time period or discipline, publishing remains a crucial step for any academic career. These authors also note that, although scientific institutions attempt to consider other artifacts, universities, grant-awarding institutions, etc. still rely heavily on bibliometric indicators. 

In the field of AI, however, some emphasis seems to be placed on these artifacts. A notable example is the ''paper with code''\footnote{https://paperswithcode.com/} platform used in this study, which indexes, sorts and ranks methods and discoveries, creating a direct link between scientific work and real-world implementation. One can also mention platforms such as Hugging face, which serve for hosting and helping users to deploy models and are widely used by academics and industrials alike, or GitHub, which is often directly referenced in the articles and redirects readers toward the code base of the project. 

Authors such as Gibney \cite{Gibney2019} or Wattanakriengkrai \cite{Wattanakriengkrai2022} mention the efforts made by the AI scientific communities to increase the availability of technical artifacts. Although this effort can be seen as a step towards more ''open science,'' it also raises questions about the nature of scientific work in the field. Put another way, have all researchers in the field shifted their focus from the paper to the code and associated repositories on platforms such as GitHub.

While questioning researchers' preferences for artifacts and the ways in which they invest time and effort in them may seem a secondary issue, it is in fact a window through which to observe a possible shift towards ''Mode 2'' of knowledge production.

In fact, if investing in the paper enables one to enact the academic credibility cycle described by Latour, authors like Alcaras \cite{alcaras2022logiciels} note that investing time and effort into code production and maintenance also generates credibility for actors within more technical or industrial arenas. 

In essence, academics and industrial researchers tend to adhere to radically different sets of rules, with the former using the code as a mere instrument to achieve publication and the latter seeing this work as a means of generating credibility (at the individual level) and financial gain (at the institutional level). Given the centrality of the industrial actors in the field, the importance attached to their research topic and the development of parallel credibility, it is more than credible to see an academic alignment and the emergence of a unified system of knowledge production.

Our study therefore aims to question this possibility: are we observing an alignment in the way artifacts are produced, published, and ultimately successful between academic and industrial actors in the field?

\newpage

\section{Data and methods}

In this paper, we utilize the "Paper with Code" database to examine the life cycle of studies disseminated through both paper and code repositories. The platform presents itself as a "free resource for researchers and practitioners to find and follow the latest state-of-the-art ML (Machine Learning) papers and code" \footnote{https://paperswithcode.com/} and features a list of open access papers associated with the technical implementation of the study.

In this context, the platform has two main advantages. Primarily, it differentiates between official and non-official implementations, enabling us to trace the evolution of a study where two concurrent artifacts are maintained by the same individual or team. Secondly, the majority of the papers linked are pre-publication versions hosted on arXiv, allowing us to track the life cycle of a study from its pre-publication stage to its subsequent publication in a journal and eventual success. \\

We began with the "Paper with Code" and filtered the 130,000 indexed entries using the classification proposed by Gargiulo et al. \cite{Gargiulo2023} This method allows us to match the different articles with their AI specialty using keyword frequency. From there, we selected four categories, encompassing both trending AI fields (classifiers, computer vision, natural language processing) where industry may have heavy involvement and declining fields (expert systems).

The objective of this selection process is to create a manageable sample that accurately reflects the current industrialized state of AI and yields the composition presented in the table below:

\begin{table}[H]
    \centering
    \begin{tabular}{|c|c|}
        \hline
        \textbf{Field of AI} & \textbf{Frequency} \\
        \hline
        Computer Vision & 2308 \\
        \hline
        Natural Language Processing & 3272 \\
        \hline
        Expert systems & 1473 \\
        \hline
        Classifiers & 2200 \\
        \hline
    \end{tabular}
    \caption{Papers with code frequency across selected fields}
    \label{tab:ai_fields_frequency}
\end{table}

To gain further insight into the publication and classify the author affiliation at the time of publication, we employed the OpenAlex API in conjunction with a manual classification of domains directly extracted from arXiv documents. To this end, we used the GROBID tool \footnote{https://grobid.readthedocs.io/en/latest/Introduction/}, which enables the extraction and parsing of scientific articles. Through this tool, we extracted the e-mail address or affiliation of the authors, depending on the availability of information, and either these were matched with existing data points or manually classified when absent.  

Furthermore, we distinguished, as suggested by Giziński et al. \cite{https://doi.org/10.48550/arxiv.2312.12881}, between articles that were solely the product of academic or industrial teams and those that involved a combination of both \footnote{For more details regarding the sampling and classification procedure, see supplementary materials \ref{sup1}}.

We can observe in the figure below the most common institutions represented in our data as well as the distribution of the studied paper and associated repositories between the distinguished groups. Here we note that our sample is mostly comprised of academic projects ($\approx79.5\%$ of the papers and associated code in the sample) and most of the industrial studies are collaboration with academic actors ($\approx 15.5 \%$ of the papers are written in collaboration between academic and industrial actors and only $\approx 5.1\% $come from purely industrial teams). 

\begin{figure}[H]
    \centering
    \includegraphics[width=1\linewidth]{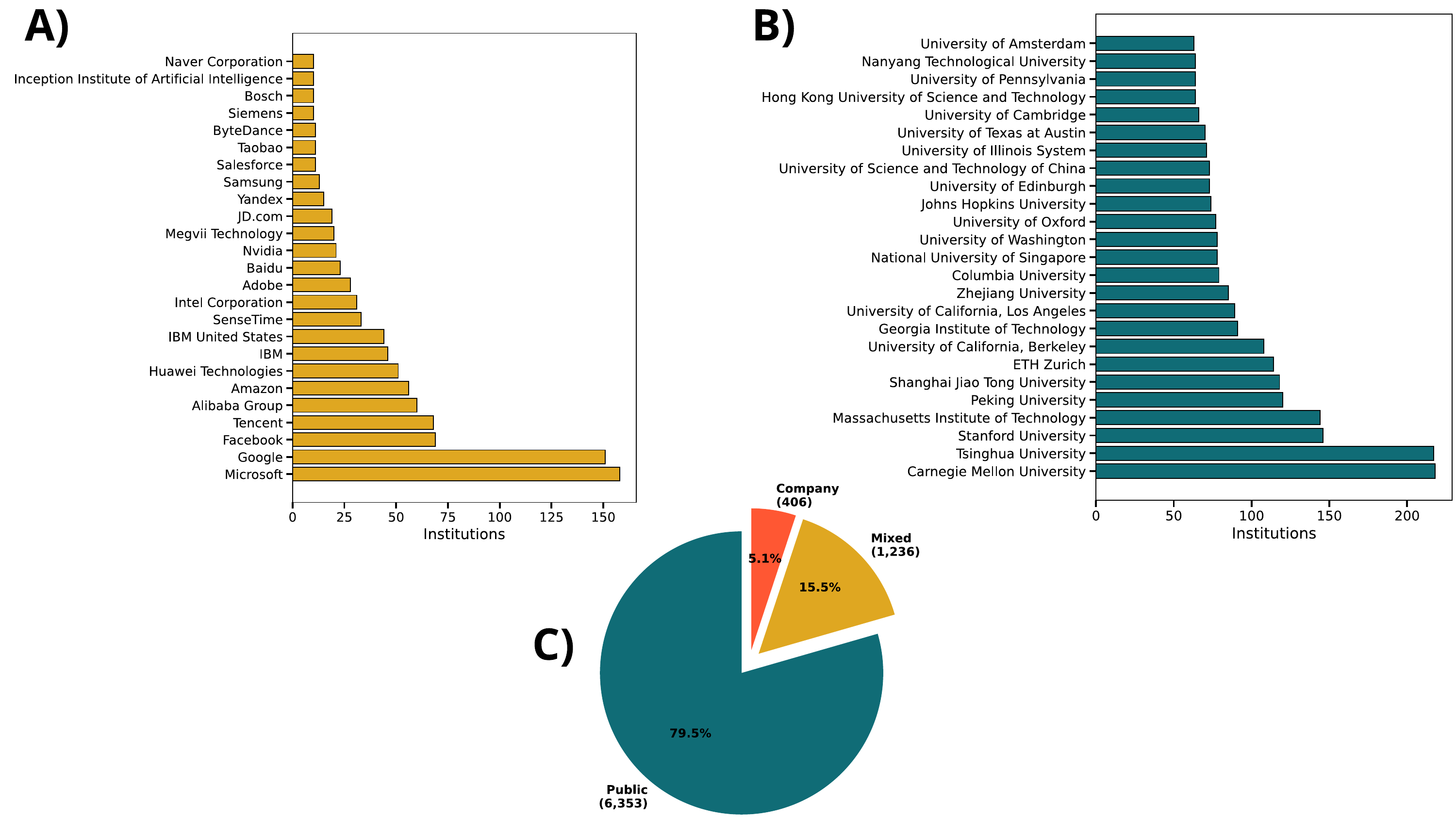}
    \caption{ \textbf{Institution type and labels in the dataset} A) Top 25 most represented industrial institution in the sample ; B) Top 25 most represented academic institution in the sample ; C) Study repartition within each group : ''Company'' represent studies conducted only by industrial actor, ''Public'' represents studies only conducted by academic or other non-profit / governmental actors and ''Mixed'' studies conducted with researchers from both groups.}
    \label{Figure0}
\end{figure}

Regarding the GitHub repositories, a comprehensive array of information was gathered via the GitHub API. GitHub provides both static data regarding the status of the repository, including its description, shared files, contributors to the project, and its current popularity and usage, as well as historical information such as the commit, stars, and issues history of the project.\\

Our analysis is conducted using a combination of scientometric indicators and metrics that have been developed for the study of open-source software projects. 
In regard to the application of scientometric analysis, our investigation is informed by the analysis conducted by Frank et al. \cite{Frank2019} and Klinger et al. \cite{Klinger2020} and their investigation about the topical diversity and success of academic and industrial articles within the field of AI. Concerning repositories, we utilized the work of Fritz \cite{fritz2010degree} \cite{fritz2014degree} and the method they developed to assess the participation of different contributors.             
In addition to concerns relating to labor intensity and distribution, we took inspiration from Gonzalez et al. \cite{Gonzalez2020} and Borges et al. \cite{Borges2018} in order to analyze the structure, usability and presentation of the repositories. 
Finally, following the works of Borges et al. \cite{Borges2018} we utilized both classic scientometric indicators as time series analysis to assess the success of both artifacts. 

\newpage

\section{Results}

\subsection{Interest and technical choices of actors}

In light of the findings presented by Frank et al. \cite{Frank2019} and Klinger et al. \cite{Klinger2020}, we conducted a quantitative analysis of the topical diversity of papers. We used OpenAlex topics to assess the diversity of academic and industrial research. We used the Shannon Entropy to determine whether one group had a greater diversity of topics covered in their studies. 

The Shannon entropy is a classic metric in information theory that measures the level of disorder in a system. In our case, it represents the uncertainty of encountering a given topic. Considering academic (\( P_a \)) and industrial papers (\( P_i \)) we extract the frequency of the topics \( T \)  for each subgroup \( Pg\) and define the entropy as follows:

\begin{equation}
    H(Pg) = -\sum_{i=1}^{n} p(T_i) \log p(T_i)
    \label{eq1}
\end{equation}

Additionally, considering that each paper \(p\) has strictly 3 topics, but that the total number of papers, ergo the size of the population \(n\) differs, we used a rarefaction process to mitigate biases related to the sample size. The process randomly selects a sample equal to the population size of the smallest subset and averages the results over a given number of iterations (1000 in our case).

\begin{figure}[H]
\begin{center}
    \centering
    \includegraphics[width=0.9\textwidth]{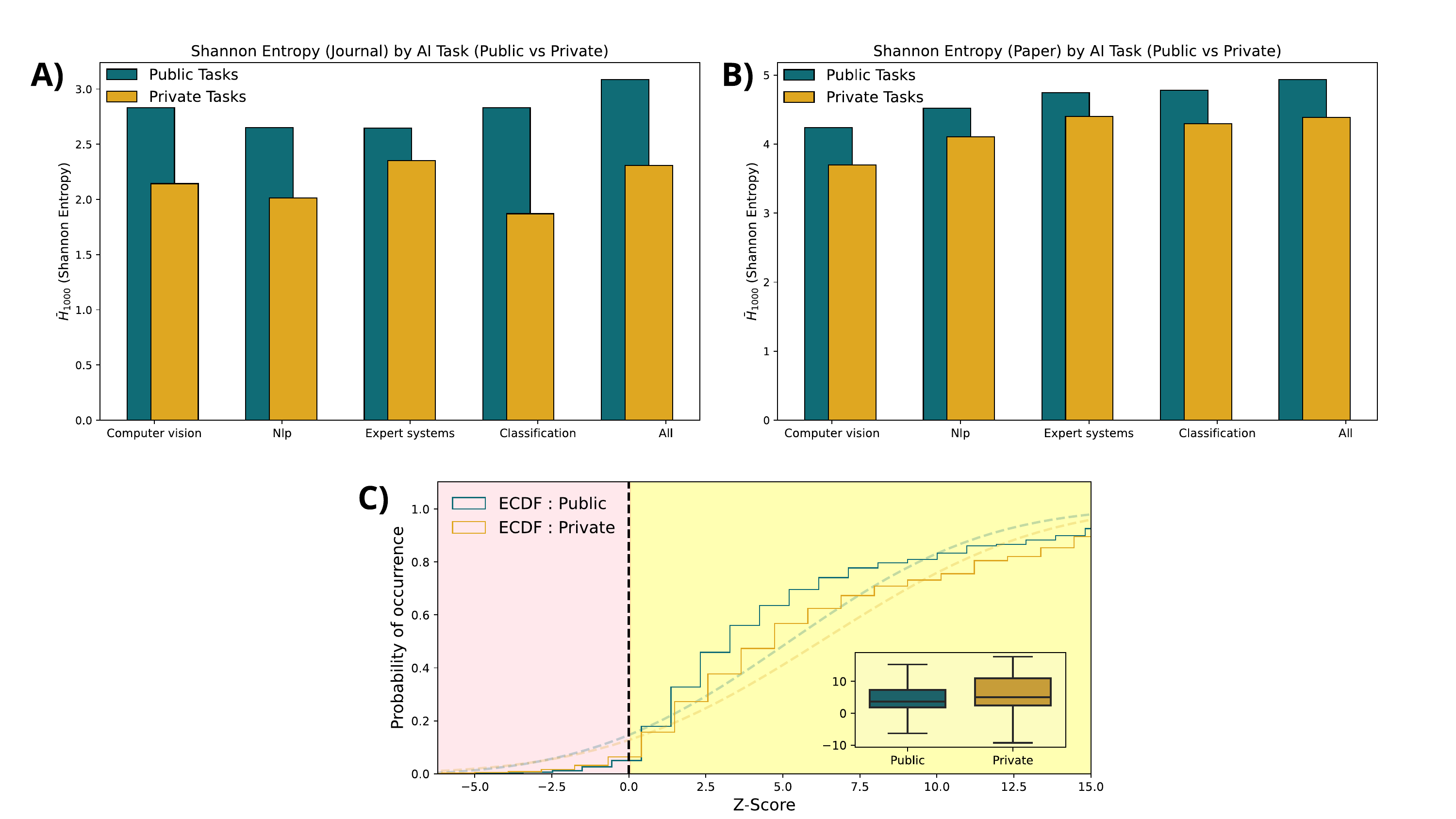}
    \caption{\textbf{Topical diversity in Academic and Industrial papers} : A) Shannon Entropy of journal's (excluding pre-publication venues) OpenAlex topics, higher value means higher topical diversity. The Shannon entropy is computed for each AI task subgroup and across groups. B) Shannon entropy for papers topic using OpenAlex Topics. C) Topic pairing z-score in academic and industrial paper highlighting the different in topic combination. Z-score computed by comparing topics pairs occurrence against a network rewiring derived baseline using Uzzi's method.}
\label{Figure2}
\end{center}
\end{figure}

As expected, for all fields of AI, we observed greater topical diversity in academia at both the journal and the article level (Fig.\ref{Figure2}.A-B). Put differently, the articles show less topical diversity and are therefore published in journals that cover fewer topics.    

In addition to single-topic diversity, we also examined the combination of topics in order to assess the propensity of researchers to explore new ideas and to incorporate an interdisciplinary approach in their work. We used Uzzi's \cite{Uzzi2013} method, which consists of creating a bipartite network between the article and the journals and iteratively rewiring this network in order to form a baseline for the computation of a Z-Score ($z = \frac{X - \mu}{\sigma} \label{eq:z_score}$). For the purposes of our study, we created a topic-topic network and proceeded with the rewiring as suggested by Uzzi. The method allows us to assess the typicality of each topic combination and, consequently, to assign a score to each paper reflecting the typicality of its concept combination. 

Our results show a clear tendency toward classic combinations for papers that include at least one industrial author (Fig.\ref{Figure2}C) suggesting that their inclusion leads to significantly less diverse scientific endeavors. 

Although not surprising, our results indicate that even within a sample of research encompassing both scientific and engineered artifacts, significant discrepancies in the researchers' interests persist. Furthermore, when purely private and mixed teams are distinguished, the results of the latter group align with those of the purely industrial teams. This suggests that the involvement of one industrial author leads to a topic alignment with industrial interest.

\begin{figure}[H]
\begin{center}
    \includegraphics[width=1\textwidth]{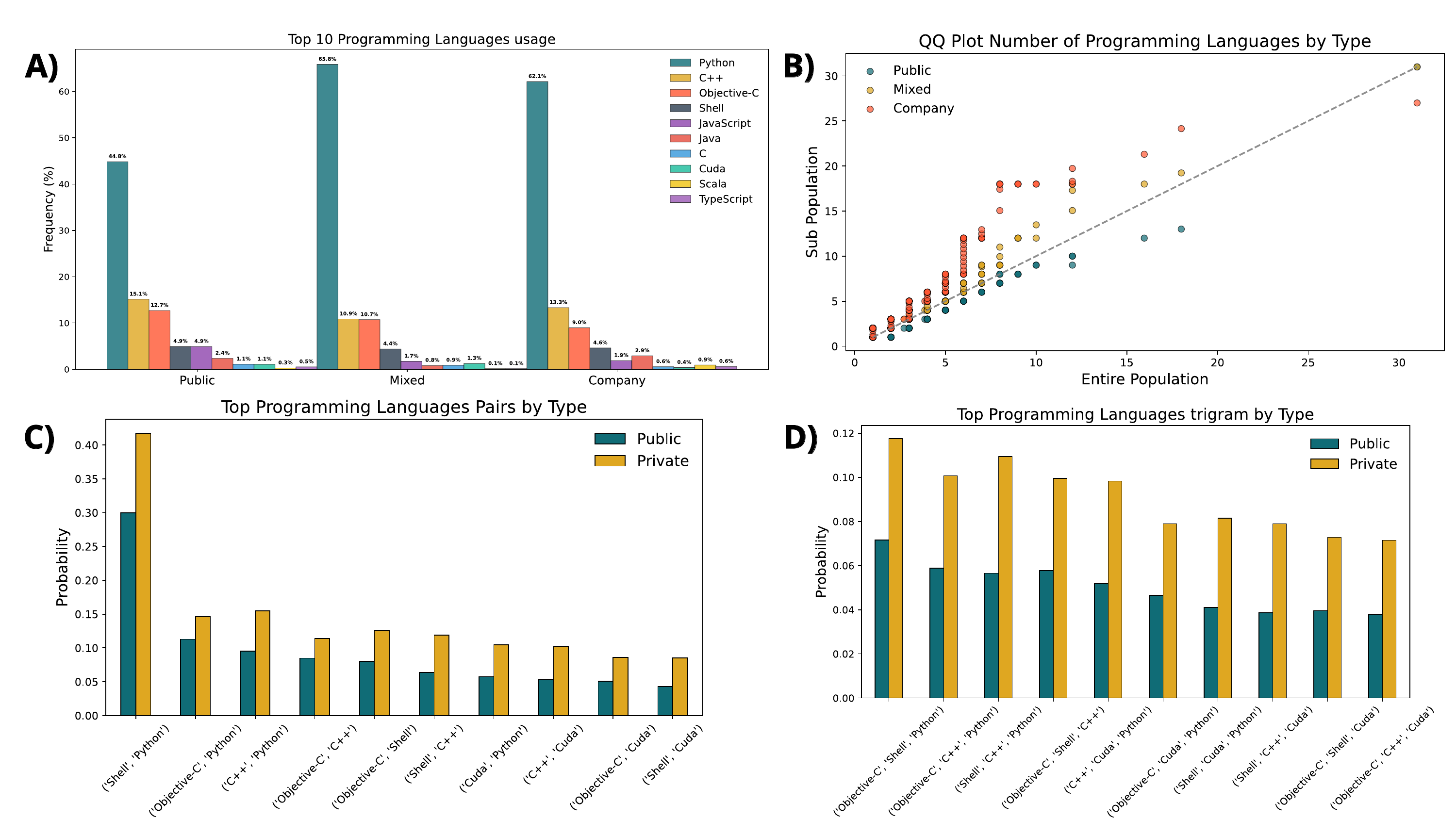}
    \caption{\textbf{Programming language in academic and industrial repositories} : A) Frequency of file programming language type for purely academic, mixed and purely industrial repositories. B) Quartile to Quartile plot for the number of programming language within the repositories for purely academic, mixed and purely industrial repositories, each points represents the group quartile and the dashed line the quartile for the entire population. C) BiGram of programming languages within academic and industrial (at least one industrial contributor) repositories. Each bar represents the probability (incidence \ref{eq2}) of encountering the pair within a repository of the group D) TriGram of programming languages within academic and industrial repositories. }
\label{Figure3}
\end{center}
\end{figure}

In addition to the differences observed at the paper level, we also note discrepancies in the repositories. A first examination of the files present in the repositories indicates that, despite an over-representation (Fig.\ref{Figure3}.A) of Python files in industrial repositories, the distribution of files within other programming languages is relatively consistent between academic, mixed, and industrial repositories. 

While similar at first glance, further investigations into the programming languages used by academics and industrials, at the repository level, demonstrate higher-level diversity in the industrial project (Fig.\ref{Figure3}.B) as well as notable differences in the programming languages used by researchers. 

To estimate the presence of a specific language within a public, private, or mixed repository, we computed the languages' prevalence within each group. To compute the prevalence within each group, we define \( P_g(L) \) as the relative presence of a language \( l \) in group \( g \) (public, private, or mixed repositories). Here, \( \text{lc}_g(l) \) represents the count of language \( l \) in group \( g \), divided by \( T_g \), the total language count in that group. This method allows for the comparison of language use across repository types as we consider the whole set of languages \( L \) to be a possibility for each group \( g \).

\begin{equation}
    P_{g}(L) = \left\{ l : \frac{\text{lc}_{g}(l)}{T_{g}} \;\middle|\; l \in L, \; g \in \{a, m, i\} \right\}
    \label{eq2}
\end{equation}

Our findings indicate significant deviations between mixed and industrial repositories in comparison to the academic ones. We first note an inversion in the importance of Python with a much lower prevalence of the language in industrial and mixed repositories ($P_a(\mathrm{Python}) \approx0.43;\;P_m(\mathrm{Python})/P_i(\mathrm{Python})\approx0.37$ ). Additionally, there is a notably higher prevalence of CUDA and Shell files within mixed repositories($P_m(\mathrm{Cuda})\approx0.05;\;P_a(\mathrm{Cuda})/P_i(\mathrm{Cuda})\approx0.03$ ). The results demonstrate the importance of industrial actors in the incorporation of costly technologies such as CUDA, and the extraction of the project from a purely Python code base \footnote{See supplementary materials \ref{sup2}}.\\

Finally, we look at the Gini index for the top 20 programming languages for each type. Ranging from 0 (perfect equality) to 1 (extreme inequality), the Gini index is a widely used synthetic indicator in sociology and economics that measures the level of inequality for a specific variable within a given population. We see a notably higher Gini index for academic repositories ($\approx 0.77$), this suggests that they could also have a less diverse code base compared to industrial ($\approx 0.64$) and mixed repositories ($\approx 0.69$).

Given the observed differences in the Gini, we turned our attention to the technical stack, specifically the combination of programming languages used in the projects. To do so we computed considered all possible pairs of programming languages for repositories and, using the same prevalence metric, we observed both the elevated diversity and the technical divide previously mentioned (Fig.\ref{Figure3}.C , Supplementary materials \ref{sup2}).\\ 

In summary, although at the file level it appears that industrials overuse Python in place of other technical options (see supplementary material \ref{sup2}), looking at the repositories as a whole reveals that they are mobilizing a wider array of technologies and are creating more complex technical stacks. Also, we note that the inclusion of certain technologies like CUDA that allow for the creation of the most efficient model or shell file to automate and facilitate the use of the code is observed much more frequently in industrial and mixed repositories. In contrast, academic codes display less complexity and diversity, often using only Python. This observation could largely be explained by the aforementioned "compute divide", but also might originate from differences in the underlying motives of the code.   

In fact, the predominance of Python as the only language in many repositories (41\% for academic repositories and 36\% for industrial ones), the simplicity of the technical stack, and the reduced effort devoted to usability can be attributed to the fact that academic code is "supplementary material" and does not aim to create complete software, but a scientific artifact that, while producing external value \cite{Brun2023}, must remain readable and follow the technical norms of its field.       

\newpage

\subsection{The life cycle of public and private studies}

The analysis conducted so far demonstrates the significant discrepancies between academics and industrials in regard to the issues addressed and the technical means to achieve their goals. While these findings already illustrate some of the differences in the researchers' practices, the life of the artifacts does not end with a manuscript and a first release on GitHub. According to Shinn \cite{Shinn2002}, industrial artifacts (e.g., technical or, in our case, the code) and academic ones (e.g., the paper) tend to have diverging paths. In fact, the former aims to become generic, or in other words, to be understood and widely used, while the latter only seeks to reach and be recognized by members of the academic community.

In order to address this question, we analyze the artifacts' path, from the first pre-publication to publication, and from the first release to the completion of the software. \\

\begin{figure}[H]
\begin{center}
    \includegraphics[width=0.9\textwidth]{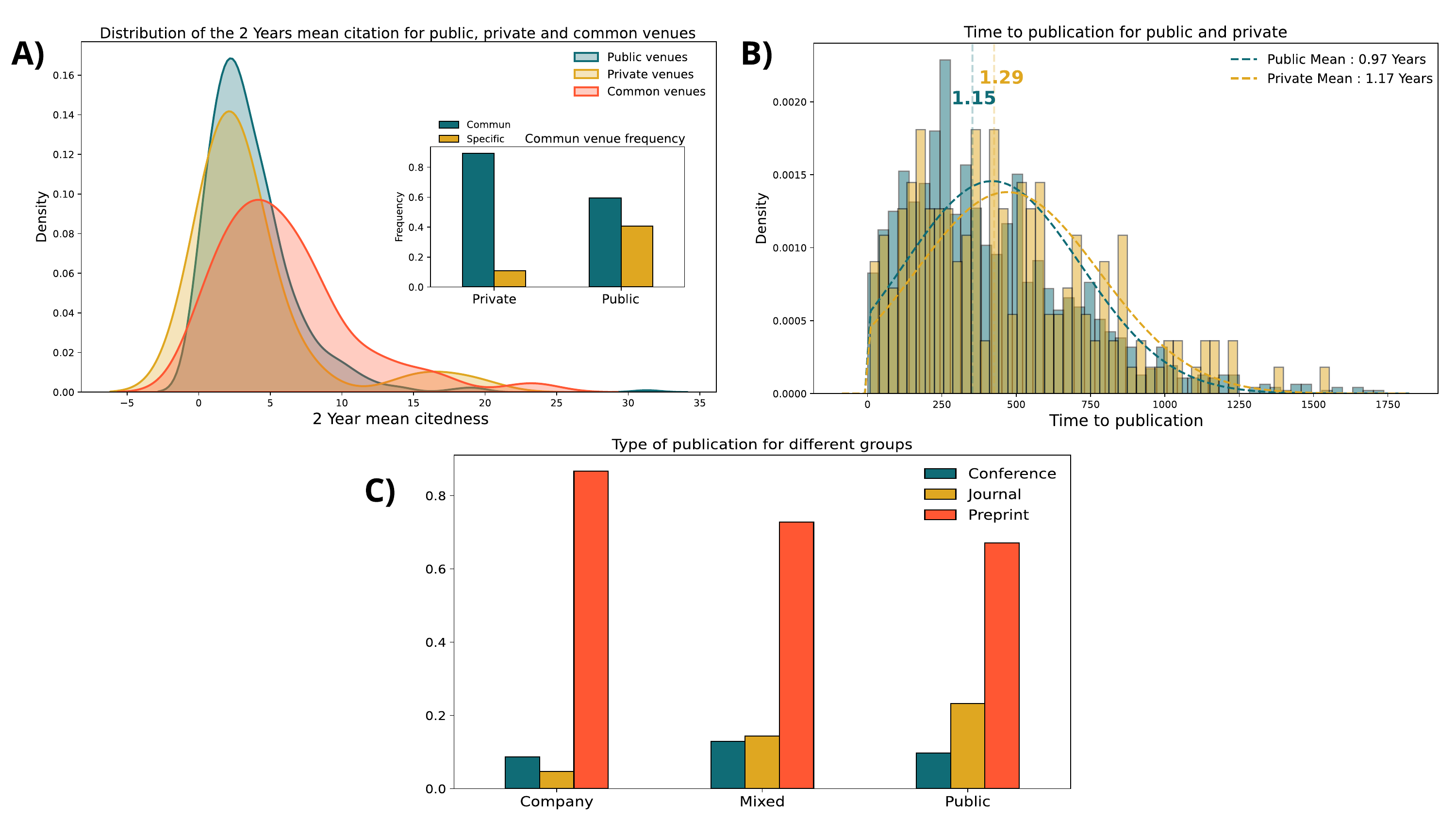}
    \caption{ \textbf{Venue choice for academic and industrial papers} : A) Density distribution function for the academic specific, industrial specific and common venues (excluding pre-publication venues). The inset shows the reparation of public and private articles within those categories. B) Histogram of time to publication for academic and industrial (at least one industrial authors). The time to publication represents the time between the first upload on arXiv and the publication of the articles. The dashed horizontal line represent a truncated normal fit and the dashed vertical line the median of the distributions. C) Bar plot of the publication status for academic, industrials or mixed teams.}

\label{Figure4}
\end{center}
\end{figure}

In regard to articles, we first note that the involvement of an industrial author results in a reduction in the number of potential venues for publication in comparison to articles published by purely academic groups. The analysis of our sample reveals that industrial authors publish their research in a more limited number of venues and primarily focus their efforts on high-impact journals (Fig.\ref{Figure4}A). In particular, the venues targeted by academic or industrial organizations specifically perform worse compared to those that the two groups have in common. However, industrial authors publish more frequently in these ''common'' venues (Fig.\ref{Figure4}A). 

This indicates that, in contrast to the "publish or perish" attitude that exists in academia \cite{Sarewitz2016PublicationFlood}, industrials may have the opportunity to be more selective in the choice of venue for their work \footnote{We also note an increased presence of predatory journals in the purely academic group, see Supplementary materials \ref{sup3}}. 

This attitude is further confirmed by the fact that within our sample industrial studies are significantly less likely to be fully published (23.2\% for academic papers and 4.7\% for industrial ones ; p < 0.0001 ; Fig.\ref{Figure4}C). Furthermore, even mixed teams are significantly less likely to have their research fully published (14.3\%), suggesting a heavy impact of industrial authors on the publication process.

Moreover, published papers including industrial authors have a longer time to publication. The mean interval between the arXiv form and the fully published article for academic articles is 0.97 years for academics and 1.17 for teams including an industrial author (T-Test p$\approx 0.03$ ; Fig.\ref{Figure4}B). The same as for the question of topics, we verified the results for the mixed situation and noted a similar trend. The average for mixed teams is closer to that of purely industrial studies than academic ones (Purely academic ($\bar{x}$ $\approx 1$) ; Mixed ($\bar{x}$ $\approx 1.2$) ; Purely industrial ($\bar{x}$ $\approx 1.3$) ANOVA test P $\approx 0.09$) confirming the importance of industrial authors in publishing decisions. 

These results show that industrials do not rely on scientific publications as much as academics do. While academics are driven by the "publish or perish" mentality, industrials have a more practical approach to publishing, only publishing certain articles, focusing on more prestigious venues and taking more time to publish their findings. Overall, it appears that academics and industrials mostly share, for the most part, their publication venues but that the latter has a more instrumental and strategic usage of the publication system. Although this strategic usage of the scientific publication system is not surprising, it already represents a significant shift from a model that segregates fundamental (e.g., academic) and technical (e.g., industrial) publication. 

Regarding repositories, we observe large deviations between academics and industrials.

\begin{figure}[H]
\begin{center}
    \includegraphics[width=1\textwidth]{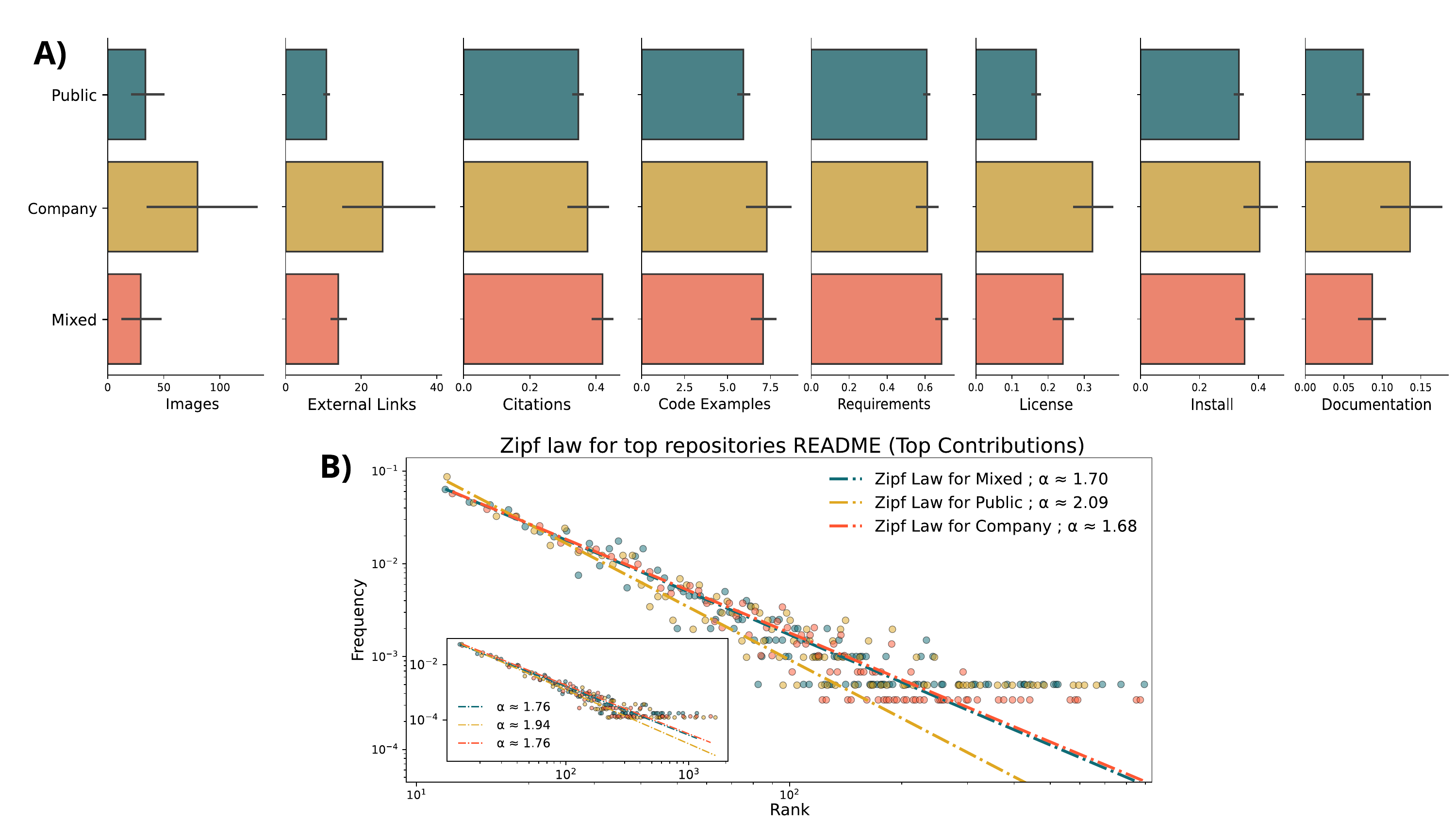}
    \caption{\textbf{Repository presentation metrics:} A) The figure shows the frequency/probability of encountering different elements within academic, mixed and industrial repositories. All repositories are considered except for the "install" variable which is filtered for repositories utilizing Python  B) Zipf law of the top repositories readme files (top 100 for each group sorted by number of commits) for academics, mixed and industrial repositories. The figures uses the word frequency across repositories readme. Inset displays the same metric with repositories filtered by number of stars (top 100). The $\alpha$ parameter relates to the lexical diversity, lower alpha indicates higher diversity.}
\label{Figure5}
\end{center}
\end{figure}

Looking at a wide array of metrics, our observations indicate that industrials appear to invest more in the presentation of the repositories, for example, by including more image files, code examples, and providing documentation more frequently (Fig.\ref{Figure5}.A). The efforts put into the repository presentation is even displayed in the lexical diversity found in the repositories' readme file's. 

To assess the lexical diversity exhibited by the repositories of the different groups, we investigated the words' frequency for the top repositories sorted by either commits or stars and computed the Zipf law associated with each distribution. The Zipf law is a classic method, derived from empirical observation, which allows us to quantify the lexical diversity in a given text. The law considers that each word frequency \( F(n) \) is linked to its rank \( n \) by a law of the form \( F(n) = \frac{\alpha}{n} \) where $\alpha$ is a constant. The constant $\alpha$ defines in this context the lexical diversity of the text with a higher parameter signaling a higher degree of inequality or lower diversity.  
Using multiple sampling strategies and sizes, we observe that purely academic repositories systematically display lower lexical diversity compared to mixed and industrial repositories (Fig.\ref{Figure5}.B , see supplementary materials \ref{sup5}). 

Despite the apparent focus on showcasing the technical artifact, we observed minimal discrepancies in repository maintenance between academic and industrial actors. In fact, for most metrics, such as the time to answer issues and the fraction of closed issues or the time between commits and the total maintenance time (see supplementary materials \ref{sup4}), we noted very little variation between groups and even observed an academic edge in some key measurements, such as the time to answer issues \footnote{We did not account for mixed repositories in those measurement due to the collection difficulties for large scale project history data.}.\\   

However, the differences reemerged when we examined the labor distribution. Due to collection difficulties and to maintain a consistent sample, we selected 100 repositories for purely academic repositories and those that have at least one industrial author.  \footnote{We filtered out repositories with less than 100 commits and more tan 2000 as to get repositories with similar level of investment. We then extracted the 100 most popular repositories (sorted by number of stars) for purely academic studies and studies including at least one industrial author.} We then computed the Shannon entropy (see equation \ref{eq2}) using in place of the topic frequency for different groups, the number of modified lines by each contributor (after the initial commit) in the repository. This approach allows us to assess the diversity in the repository authorship or, in other words, the distribution of labor within the project. 

\begin{figure}[H]
\begin{center}
    \includegraphics[width=1\textwidth]{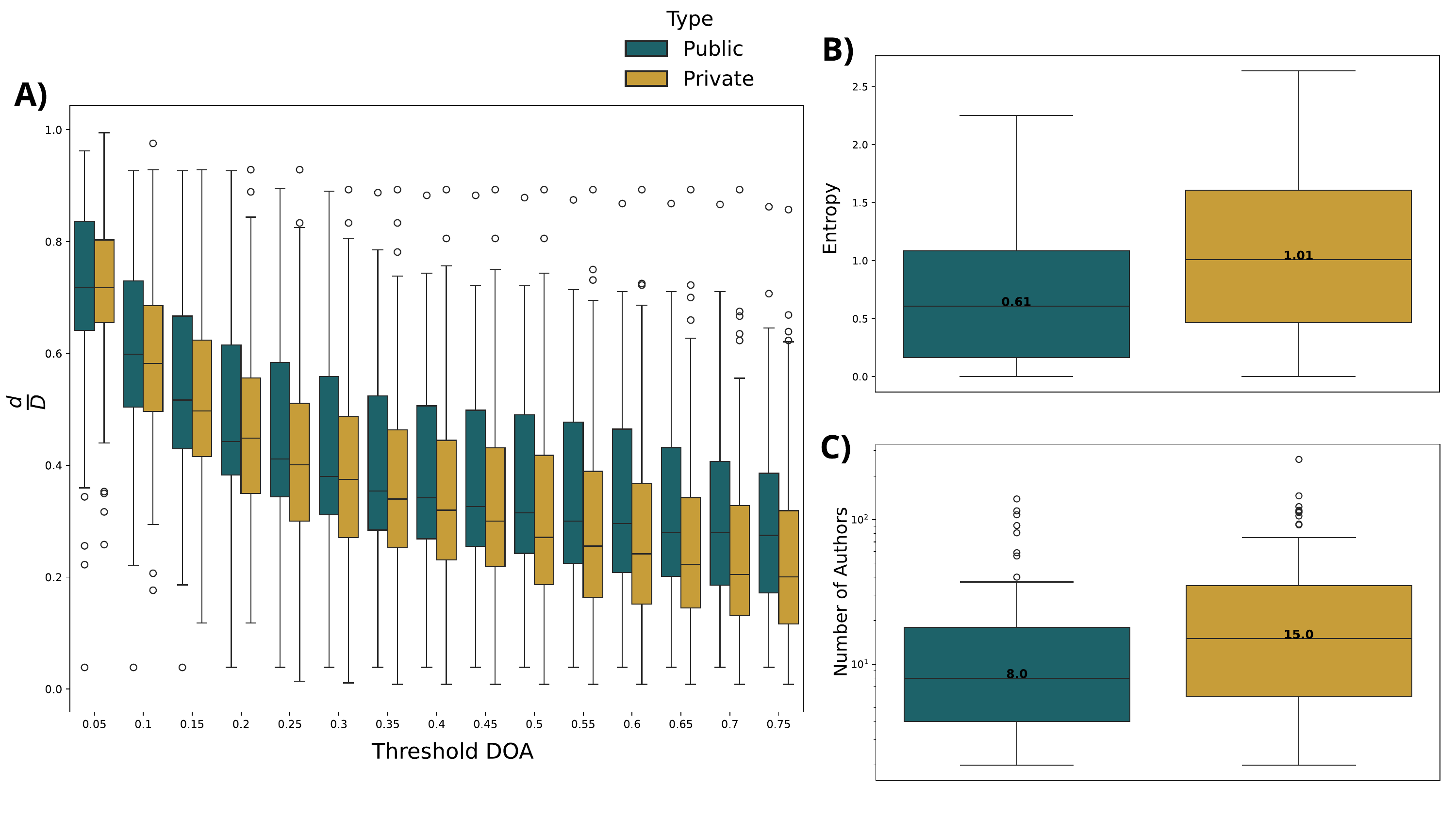}
    \caption{ \textbf{Repository maintenance by academic and industrial team} : A) The figure shows, for each file in industrial and academic repositories, the fraction of contributors considered as "author" according to the threshold. B) Entropy of line modification for academic and industrial repositories. C) Count of authors (DOA threshold = 0.75) within academic and industrial repositories.}

\label{Figure6}
\end{center}
\end{figure}

We observe a significantly higher entropy for industrial repositories which we can interpret as a signal that the work on the code is distributed more evenly among the contributors to the project (Fig.\ref{Figure6}). However, such results are lacking as they aggregate all contributions that obscure differences at the file level. To confirm this result, we borrowed the DegreeOfAuthorship \ref{eq3} (DOA \cite{fritz2010degree} \cite{fritz2014degree}) metric from Fritz et al. and its implementation by Avelino et al \cite{Avelino2019}.

\newpage

For each contributor \( d \) and source element \( f \), the DOA is calculated by the combination of weighted factors: \( FA \) indicating if the contributor \( d \) is the initial author, \( DL \) the number of lines changed by the contributor \( d \), and \( AC \) for the number of changes accepted by others, with logarithmic decay.\\

\begin{equation}
    DOA_A(d, f) = 3.293 + 1.098 \cdot FA + 0.164 \cdot DL - 0.321 \cdot \log(1 + AC)
    \label{eq3}
\end{equation}

In order to allow comparison at the repository level, we used the same normalization technique as Avelino \cite{Avelino2019}. 

\begin{equation}
    DOA_N(d, f) = \frac{DOA_A(d, f)}{\max \{ DOA_A(d', f) \mid d' \in \text{changed}(f) \}}
    \label{eq4}
\end{equation}

This measure provides to each author \( d \) a normalized score, ranging from 0 to 1 which represents their investment in the writing of a file \( f \). Using different thresholds, we can consider the lowest investment score that qualifies a contributor as an author of a file \( f \). After computing the normalized DOA for each contributor / file pair in our repository subset, we considered different thresholds and chose to stop at a normalized score of 0.75 to consider a contributor as author (results for higher thresholds are consistent with these observations). This value was chosen by Avelino \& colleagues \cite{Avelino2019} in their study on highly collaborative open source projects and appeared to be a reasonable upper bound.\\

Our measurements corroborate our initial results; industrial repositories have a lower proportion of contributors designated as authors for each file. Additionally, using the same 0.75 threshold and looking at the raw count of authors in each repository ($\left| \{ d \in D \mid  \mathrm {DOA}(d) > 0.75 \} \right|$), we note that industrial repositories have simply more people involved in each project (Academic $\bar{x}$ $\approx 16.9$ \& Industrial  $\bar{x}$ $\approx 29.2$ ; T-test p $\approx 0.009$).\\

\newpage

These results, taken together, illustrate a real academic / industrial divide when it comes to research practice. Academics tend to publish their results in peer review journals faster and at a higher frequency and are more amenable to having their work published in lower-impact journals. This tendency contrasts with industrials, who publish at a slower pace and primarily focus on high impact venues.

With regard to the repository, although maintenance practices are similar, industrials put more effort into their project presentation and usability and maintain these complex projects with a higher number of people and a better partition of labor.  

\subsection{Public attention to the different artifacts}

\begin{figure}[H]
\begin{center}
    \includegraphics[width=1\textwidth]{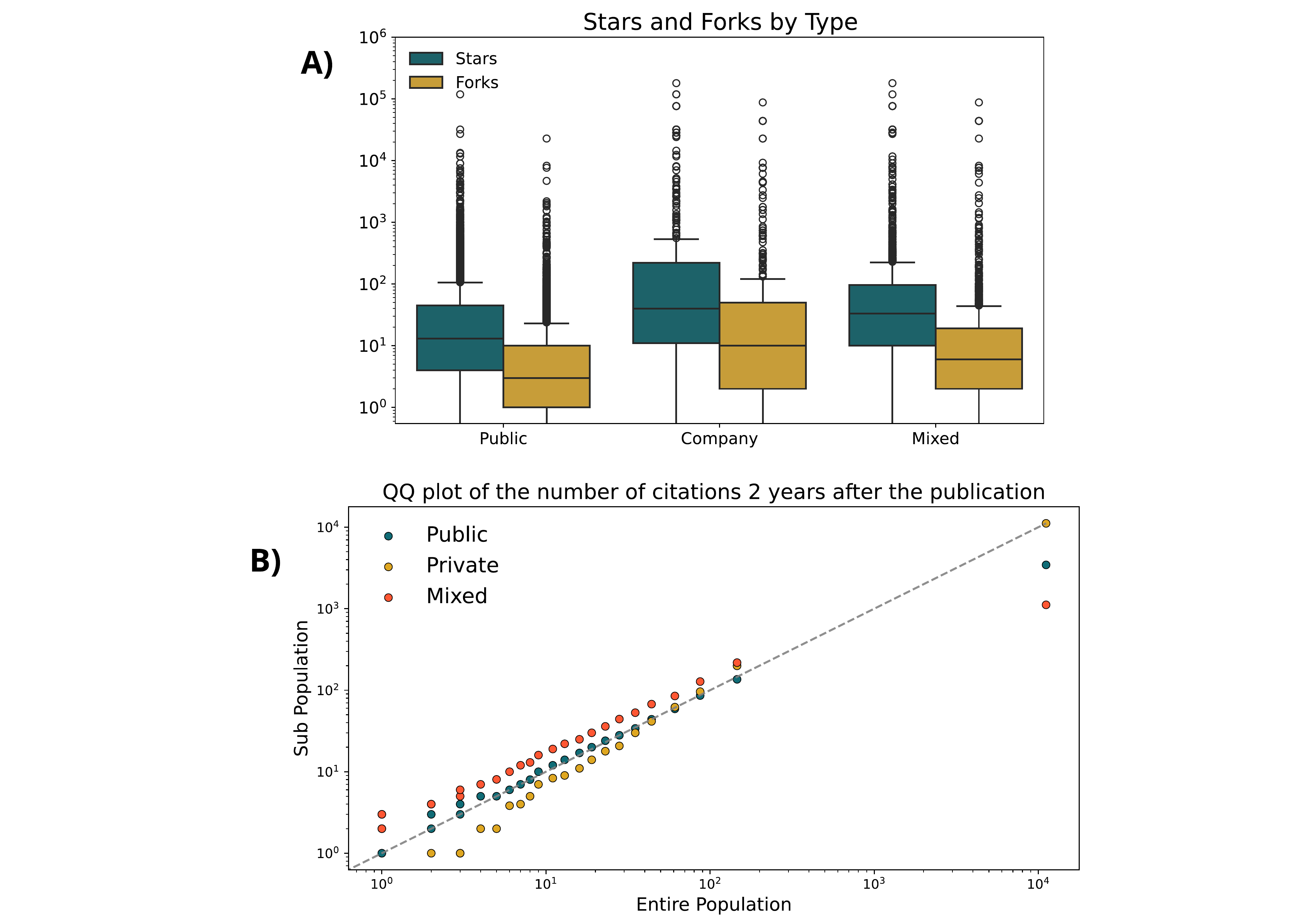}
    \caption{ \textbf{Artifact performance} : A) Box plot for the number of stars and forks for Github repositories linked to purely public, purely private and mixed studies. B) The figure shows a quantile–quantile plot for the purely industrials, purely academics and mixed teams for the numbers of citation after 2 years. The gray dashed line shows the entire population quantile and points the quantile for each sub group.}

\label{Figure7}
\end{center}
\end{figure}

The final stage of any artifact life cycle is reaching a public. Using citations for scientific articles and GitHub's popularity metrics such as stars and forks, we analyzed the relative success of each group in each arena. 
The figure below (Figure.\ref{Figure7} shows the performance of papers and repositories according to different metrics. For repositories, we considered performance in stars and forks, which help to assess public interest on Github, and for papers we mobilized citation as a proxy for academic attention.
Here we note that a mixed composition is always advantageous. However, purely academic papers tend to perform better in terms of citations compared to purely industrial ones. Conversely, purely industrial papers demonstrate significantly higher performance than academic ones in GitHub's metric, and similarly with mixed ones.

\newpage
While other studies demonstrate comparable trends, it is worth noting that within our specific sample, this discrepancy emerges. In other words, the publication of code, its indexing on a global platform, and its maintenance do not offset the gap observed in previous studies. 

If the disparity in terms of visibility is evident, we also observe discrepancies in the accumulation pattern of stars and citations. Following the methodology proposed by Borges \cite{Borges2018}, we investigated the growth patterns of stars and citations using a time series clustering method.

As both citations and GitHub stars are discrete occurrences, our initial list of events was transformed into a cumulative weekly time series. For this transformation, only events occurring during the first two years of the artifact's life were considered, with values normalized based on this two-year window. In other words, given \( s \) the sequence of events that occur between the time \( t_0 \) and \( t_0 + 2y \), we can consider the total number of events during this time window as \( s(t_{y2}) \) and compute the fraction of events \( F \) after \( n \) weeks as: \( F_n = \frac{s(t_n)}{s(t_{y2})} \). Finally, we filtered articles and repositories with a low event count, thus enabling a meaningful comparison of the cumulative trend and avoiding clusters composed of stationary time series.\\

To perform the clustering, we utilized an implementation of the K-Means algorithm \cite{JMLR:v21:20-091} for time series data. The optimal number of clusters was then determined using the $\beta \text{cv}$ heuristic. The $\beta \text{cv}$ is defined as the ratio of variation intra- and inter-cluster \ref{eq4}. 

Based on this metric and a classic elbow method\footnote{see supplementary material \ref{sup6}}, we have determined that the optimal number of clusters $\mathrm{k}$ for the citation and the star time series is four. 

\begin{equation}
    \beta_{\text{CV}} = \frac{\frac{1}{K} \sum_{k=1}^{K} \frac{1}{|C_k|} \sum_{x_i, x_j \in C_k} d(x_i, x_j)}{\frac{1}{K(K-1)} \sum_{k=1}^{K} \sum_{l \neq k} d(C_k, C_l)}
    \label{eq5}
\end{equation}

\begin{figure}[H]
\begin{center}
    \includegraphics[width=1.1\textwidth]{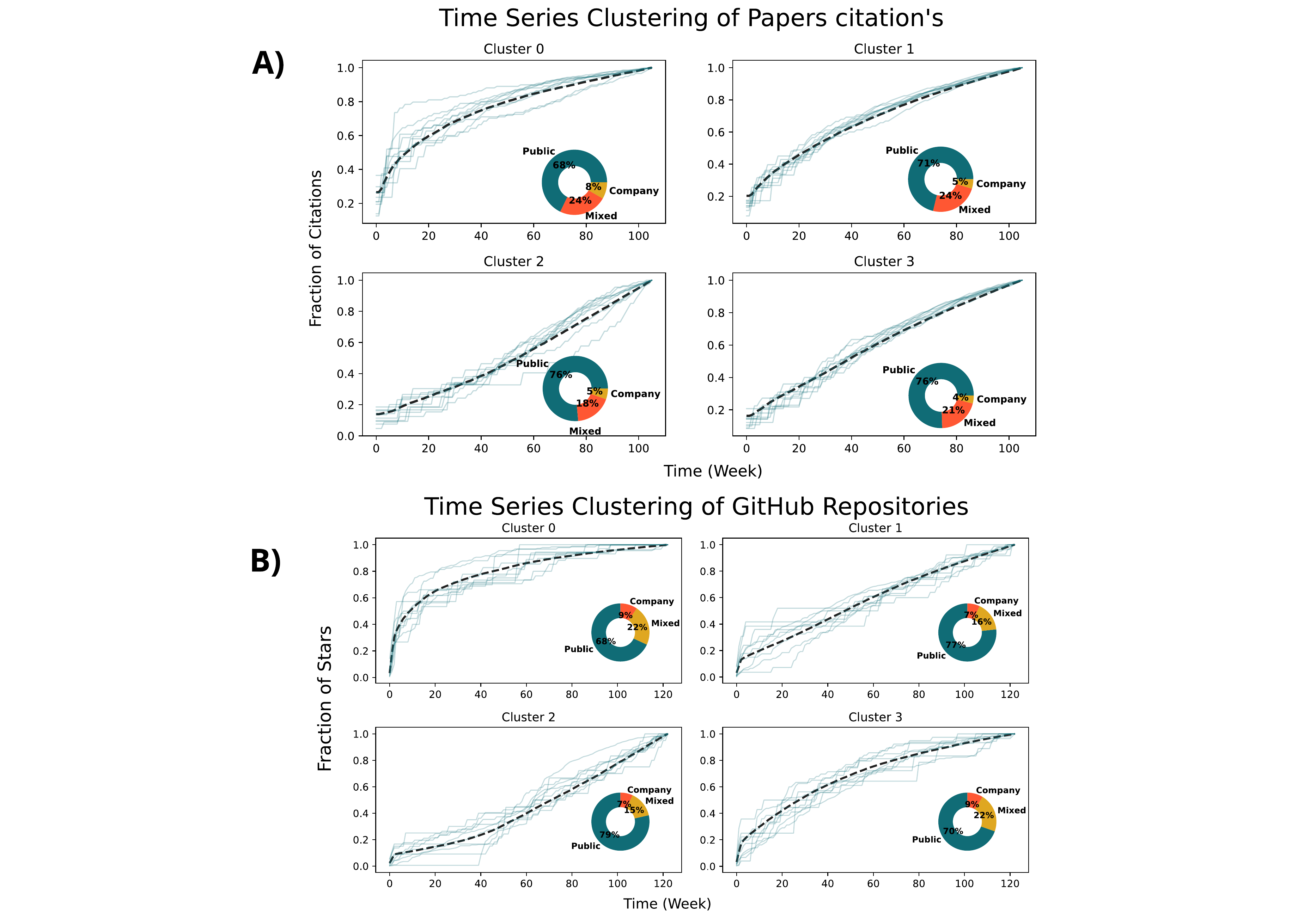}
    \caption{ \textbf{Artifact performance over time} : A) K-Means clustering of the cumulative time series of stars for a sample of repositories. The figure is showing each cluster with the dashed grey line being the general trend of the cluster and the transparent green line the time series considered (sample) in the cluster. The pie charts show the cluster composition by the different groups. B) K-Means clustering of the cumulative time series of citations for a sample of papers.}

\label{Figure8}
\end{center}
\end{figure}

In regard to GitHub stars, we note that clusters exhibiting supra-linear growth of stars are disproportionately represented by industrial and mixed repositories. In contrast, academic repositories are predominantly linked to clusters with linear or sub-linear growth (Cluster 0 : 32\% industrials ; Cluster 4 : 30\% industrials against 23\% and 21\% for cluster 1/2 ; Figure.\ref{Figure8}).
We observe a similar trend with regard to citations. Here, we find that industrial and mixed papers are overrepresented in the cluster with exponential growth (Cluster 2 : 32\% industrials ; Cluster 1 : 29\% industrials against 24\% for cluster 0/3 Figure.\ref{Figure8}). In general, it seems that teams including at least one industrial actor tend to achieve success more quickly for both artifacts.

The results indicate that an industrial contribution to the study increases the likelihood of attracting the attention of the community in a relatively short period of time. This situation may be attributed to a number of factors, including an enhanced capacity to promote the artifact or the observation that industrial research yields more pioneering and immediately applicable results. 

Despite this initial advantage, a clear segregation in the ultimate success of the artifact is still noted. Purely academic teams tend to outperform industrial ones with regard to citations. In contrast, purely industrial teams accumulate a greater number of GitHub stars. This suggests that industrial productions have greater and longer-lasting technical value compared to academic ones that focus on the heuristic benefits of their productions. 

Finally, we observed that the mixed composition represents the most advantageous scenario for all parties involved, or at the very least, the one that yields the most successful outcomes on both Scholar and GitHub.

\newpage
\section{Discussion}

The objective of this study was to observe the potential alignment between academic research and industrial practices in the field of artificial intelligence (AI) research and investigate the possible formation of a unified ''Mode'' of knowledge production. 

Shinn's initial assessments \cite{Shinn2002} suggested that although the ties between academia and other actors were intensifying, the concrete research output or artifact remained distinct. In other words, industrial actors aim to create generic technologies with broad impact, while academics seek to address only their community and its underlying interests and goals.   

Given the current state of techno-scientific development of AI, and in particular the industrial dominance observed by many researchers, we hypothesized that a new synchronicity between academia and industry could be observed. Such an alignment would provide strong evidence for the existence of a "Mode 2" of knowledge production \cite{Gibbons1994-qi}. 

In order to assess these potential alignments, we propose a comparative study between the Github repositories and research papers. A review of these artifacts allowed us to assess the underlying motives of the researcher and the potential change in the norms guiding their work\cite{grossetti2000science}\cite{Shinn2002}\cite{Raimbault2023}. 

Our aim was to assess whether and to what extent the industrial domination in the field of AI, as well as the ever more frequent switch between academia and industry, resulted in the adoption of industrial interests, norms, and practices by academics. 

The results of this study show that although both actors write and share both code and paper, their practices and the norms guiding them differ greatly. On the one hand, academics appear to prioritize the rapid publication of articles, a practice that is often driven by the imperative of "publish or perish." While this necessity leads academic authors to publish in lower-impact journals, it also appears to encourage the exploration of more novel ideas.
In examining the repositories, we observe that the academic projects tend to be significantly simpler in both their technical and presentation aspects. Additionally, although these repositories are typically well-maintained, they are often managed by only a handful of individuals. 

On the other hand, industry-specific publications tend to focus on a narrower and potentially more applied set of issues \cite{Klinger2020}. In addition, unlike academics, teams comprising industrial authors appear to publish less frequently, focus on high-impact journals, and devote more time to the publication process. Overall, industrial authors appear to have their own set of issues and only submit their work to top journals if it aligns with the publication's criteria.
In contrast, industrial actors appear to invest more effort in the repository. Despite initial observations of minimal discrepancies in maintenance, industrial repositories exhibit a higher level of technical complexity, superior presentation, and usability, and consequently require a larger and more specialized developer team for maintenance.  

However, some elements also point to a certain degree of convergence. The fact that industrials share most of their publishing venues with academics contrasts with previous results \cite{Shinn2002}. Also, as mentioned above, the increased investment of academics in repository maintenance shows the importance given to the technical artifact.  

In sum, while we can observe some level of alignment by academics, it seems that even in the field of AI, the academic code remains a "supplementary" material.

The publication bias that many academics exhibit can be easily explained by reference to Latour and Woolgar's cycle of scientific credibility \cite{latour2013laboratory}. As mentioned in our introduction, the concept simply tells us that the act of publishing, and the subsequent credibility it brings, is crucial to staying in academia. Although there are other sources of credibility, namely a technical one that can be acquired through GitHub \cite{alcaras2022logiciels}, the preferred arena for academics remains the traditional Latourian one.

This situation can be understood using Brun \cite{Brun2023} work which demonstrated that, even in disciplines with significant industrial ties, artifacts that do not align with established academic standards are frequently overlooked by academic institutions. While researchers in these fields may perceive such outputs as necessary for their own advancement as well as for the advancement of the field, other stakeholders, including universities, grant-awarding bodies, and individuals from other domains, tend to disregard these contributions. 

In general, despite the growing influence of industry within this field, academic norms and practices maintain relevance, and many academics still practice "normal science" \cite{Funtowicz1993} \cite{Brggemann2020}.

The Industry, on the other hand, seems to mobilize the artifact in a relatively instrumental way. These results fall in line with those of authors such as Baruffaldi \cite{Baruffaldi2020}, who point to the benefits drawn by industrials engaging in scientific research. 

In fact, familiarity with the scientific norms, long-term investment in scientific communities, as well as physical proximity (for example, in conferences) is crucial to enable a good flow of knowledge between academics and industrials. This idea seems to be supported by the low publication rate as well as the choice of venues for their work as well as by the over-representation of certain institutions in the literature (As noted in the introduction ; Figure.\ref{Figure0}) 

We can understand this peculiar practice of publication as a strategic way to maintain a connection to the scientific world, leverage the reputation and prestige of scientific institutions, foster knowledge exchange, and allow for collaborations. In regard to this topic of collaborations, we observe that although purely academic and industrial productions remain differentiated, the situation of mixed teams presents an interesting case.

The inclusion of an industrial author on the research team appears to greatly influence the direction of the project, with a tendency towards industrial-oriented choices and practices. Additionally, we observed that mixed teams find greater success across the board.    
In light of these observations and the results mentioned above, we can understand the situation using Moore \& colleagues concept of "asymmetrical convergence" \cite{Frickel2006-pk}. This concept refers to the idea that, while norms and practices may converge, this rapprochement primarily benefits industrial actors. 

The ability to influence the research topic, the publication process, and the design of the technical artifact suggests that, in exchange for additional resources and greater success, academics must concede some of their own norms and practices. Moreover, even when academics manage to create artifacts that they can value in their field, it still benefits industrial actors, as they can also find legitimacy through them. 

Ultimately, there seems to be a degree of alignment between academic and industrial teams, even if it does not directly affect purely academic teams. Such observations do not allow us to say that "Mode 2" has become a reality, but they do show that industry is becoming a key actor in the development of AI science and should prompt us to call for the strengthening of academia.

\section{Conclusions}

Our study used multiple data sources to assess the impact of this new academic-industrial proximity and the dominance of the latter in the field. Although we noted some similarities, the choices, investment, and success of the scientific and technical artifact remained highly differentiated between academic and industrial teams. This suggests that even if at an individual level proximity can exist (as seen with the ever more frequent switch in between academia and industry), at the institutional level very distinct norms remain. 

However, we have found that a mixed situation (having a study conducted by academic and industrial researchers) offers a distinct advantage. Although academics and industrials find relatively differentiated success, mixed teams are successful across both artifacts. Additionally, we observe some level of alignment of the mixed team with the industrial practices, suggesting that if there is convergence, it is passing through those mixed teams rather than through pure academic or industrial studies. 

This dynamic is not new and has already been conceptualized by authors such as Moore \& colleagues as an "asymmetrical convergence". The concept describes a situation where convergence is observed with unequal benefits, typically for the industry. 

While observing such dynamics is not surprising, their effects are usually constrained. In most disciplines, the industrial influence is localized. However, in the case of AI research, industrial requests appear to affect the integrity of the field. In fact, the narrowing in topics and methods that authors such as Klinger \cite{Klinger2020} or Giziski \cite{https://doi.org/10.48550/arxiv.2312.12881} document is probably fueled by this growing influence of industrial interest and the increased academic necessity for their resources.  

In conclusion, we can only, as Giziński et al., call for a strengthening of academic AI research to ensure that the field can continue to explore new ideas, methods, and invest in artifacts with long-lasting heuristic value. 
\newpage

\bibliographystyle{acm.bst}
\bibliography{sample} 

\begin{thebibliography}{10}

\bibitem{ahmed2020democratization}
{\sc Ahmed, N., and Wahed, M.}
\newblock The de-democratization of ai: Deep learning and the compute divide in artificial intelligence research.
\newblock {\em arXiv preprint arXiv:2010.15581\/} (2020).

\bibitem{Ahmed2023}
{\sc Ahmed, N., Wahed, M., and Thompson, N.~C.}
\newblock The growing influence of industry in ai research.
\newblock {\em Science 379}, 6635 (Mar. 2023), 884–886.

\bibitem{alcaras2022logiciels}
{\sc Alcaras, G.}
\newblock {\em Des logiciels libres au contr{\^o}le du code: l'industrialisation de l'{\'e}criture informatique}.
\newblock PhD thesis, Paris, EHESS, 2022.

\bibitem{Avelino2019}
{\sc Avelino, G., Passos, L., Hora, A., and Valente, M.~T.}
\newblock Measuring and analyzing code authorship in 1 + 118 open source projects.
\newblock {\em Science of Computer Programming 176\/} (May 2019), 14–32.

\bibitem{Barrier2014}
{\sc Barrier, J.}
\newblock Partenaires particuliers : financements sur projet et travail relationnel dans les réseaux de collaboration science-industrie.
\newblock {\em Genèses n° 94}, 1 (May 2014), 55–80.

\bibitem{Baruffaldi2020}
{\sc Baruffaldi, S., and P\"{o}ge, F.}
\newblock A firm scientific community: Industry participation and knowledge diffusion.
\newblock {\em SSRN Electronic Journal\/} (2020).

\bibitem{Borges2018}
{\sc Borges, H., and Tulio~Valente, M.}
\newblock What’s in a github star? understanding repository starring practices in a social coding platform.
\newblock {\em Journal of Systems and Software 146\/} (Dec. 2018), 112–129.

\bibitem{Brggemann2020}
{\sc Br\"{u}ggemann, M., L\"{o}rcher, I., and Walter, S.}
\newblock Post-normal science communication: exploring the blurring boundaries of science and journalism.
\newblock {\em Journal of Science Communication 19}, 03 (June 2020), A02.

\bibitem{Brun2023}
{\sc Brun, V.}
\newblock "les brevets sont à peine au rang d’une publication" : Projets de valorisation et cycle de crédibilité au cnrs.
\newblock {\em Revue d’anthropologie des connaissances 17}, 2 (May 2023).

\bibitem{Brunet2012}
{\sc Brunet, P., and Dubois, M.}
\newblock Cellules souches et technoscience : sociologie de l’émergence et de la régulation d’un domaine de recherche biomédicale en france.
\newblock {\em Revue fran\c{c}aise de sociologie Vol. 53}, 3 (Sept. 2012), 391–428.

\bibitem{Cardon2018}
{\sc Cardon, D., Cointet, J.-P., and Mazières, A.}
\newblock La revanche des neurones: L’invention des machines inductives et la controverse de l’intelligence artificielle.
\newblock {\em Réseaux n° 211}, 5 (Nov. 2018), 173–220.

\bibitem{duPlessis2008}
{\sc du~Plessis, M.}
\newblock The strategic drivers and objectives of communities of practice as vehicles for knowledge management in small and medium enterprises.
\newblock {\em International Journal of Information Management 28}, 1 (Feb. 2008), 61–67.

\bibitem{Frber2023}
{\sc F\"{a}rber, M., and Tampakis, L.}
\newblock Analyzing the impact of companies on ai research based on publications.
\newblock {\em Scientometrics 129}, 1 (Nov. 2023), 31–63.

\bibitem{Felt2016}
{\sc Felt, U.}
\newblock {\em Of Timescapes and Knowledgescapes}.
\newblock Oxford University Press, Dec. 2016, p.~129–148.

\bibitem{Frank2019}
{\sc Frank, M.~R., Wang, D., Cebrian, M., and Rahwan, I.}
\newblock The evolution of citation graphs in artificial intelligence research.
\newblock {\em Nature Machine Intelligence 1}, 2 (Feb. 2019), 79–85.

\bibitem{Frickel2006-pk}
{\sc Frickel, S., and Moore, K.}, Eds.
\newblock {\em The new political sociology of science}.
\newblock Science \& Technology in Society. University of Wisconsin Press, Madison, WI, Mar. 2006.

\bibitem{fritz2014degree}
{\sc Fritz, T., Murphy, G.~C., Murphy-Hill, E., Ou, J., and Hill, E.}
\newblock Degree-of-knowledge: Modeling a developer's knowledge of code.
\newblock {\em ACM Transactions on Software Engineering and Methodology (TOSEM) 23}, 2 (2014), 1--42.

\bibitem{fritz2010degree}
{\sc Fritz, T., Ou, J., Murphy, G.~C., and Murphy-Hill, E.}
\newblock A degree-of-knowledge model to capture source code familiarity.
\newblock In {\em Proceedings of the 32nd ACM/IEEE International Conference on Software Engineering-Volume 1\/} (2010), pp.~385--394.

\bibitem{Funtowicz1993}
{\sc Funtowicz, S.~O., and Ravetz, J.~R.}
\newblock Science for the post-normal age.
\newblock {\em Futures 25}, 7 (Sept. 1993), 739–755.

\bibitem{Gargiulo2023}
{\sc Gargiulo, F., Fontaine, S., Dubois, M., and Tubaro, P.}
\newblock A meso-scale cartography of the ai ecosystem.
\newblock {\em Quantitative Science Studies 4}, 3 (2023), 574–593.

\bibitem{Gelles2024}
{\sc Gelles, R., Kinoshita, V., Musser, M., and Dunham, J.}
\newblock Resource democratization: Is compute the binding constraint on ai research?
\newblock {\em Proceedings of the AAAI Conference on Artificial Intelligence 38}, 18 (Mar. 2024), 19840–19848.

\bibitem{Gibbons1994-qi}
{\sc Gibbons, M., Limoges, C., Nowotny, H., Schwartzman, S., Scott, P., and Trow, M.}
\newblock {\em The new production of knowledge}.
\newblock SAGE Publications, London, England, July 1994.

\bibitem{Gibney2019}
{\sc Gibney, E.}
\newblock This ai researcher is trying to ward off a reproducibility crisis.
\newblock {\em Nature 577}, 7788 (Dec. 2019), 14–14.

\bibitem{https://doi.org/10.48550/arxiv.2312.12881}
{\sc Giziński, S., Kaczyńska, P., Ruczyński, H., Wiśnios, E., Pieliński, B., Biecek, P., and Sienkiewicz, J.}
\newblock Big tech influence over ai research revisited: memetic analysis of attribution of ideas to affiliation, 2023.

\bibitem{Gonzalez2020}
{\sc Gonzalez, D., Zimmermann, T., and Nagappan, N.}
\newblock The state of the ml-universe: 10 years of artificial intelligence ; machine learning software development on github.
\newblock In {\em Proceedings of the 17th International Conference on Mining Software Repositories\/} (June 2020), MSR ’20, ACM.

\bibitem{grossetti2000science}
{\sc Grossetti, M., and Detrez, C.}
\newblock Science d'ing{\'e}nieurs et sciences pour l'ing{\'e}nieur: l'exemple du g{\'e}nie chimique.
\newblock {\em Sciences de la soci{\'e}t{\'e}: Les cahiers du LERASS\/} (2000), pp--63.

\bibitem{Hagendorff2021}
{\sc Hagendorff, T., and Meding, K.}
\newblock Ethical considerations and statistical analysis of industry involvement in machine learning research.
\newblock {\em AI \& SOCIETY 38}, 1 (Sept. 2021), 35–45.

\bibitem{Hessels2019}
{\sc Hessels, L.~K., Franssen, T., Scholten, W., and de~Rijcke, S.}
\newblock Variation in valuation: How research groups accumulate credibility in four epistemic cultures.
\newblock {\em Minerva 57}, 2 (Jan. 2019), 127–149.

\bibitem{jacobides2021evolutionary}
{\sc Jacobides, M.~G., Brusoni, S., and Candelon, F.}
\newblock The evolutionary dynamics of the artificial intelligence ecosystem.
\newblock {\em Strategy Science 6}, 4 (2021), 412--435.

\bibitem{Joly2019}
{\sc Joly, P.-B.}
\newblock {\em Reimagining Innovation}.
\newblock Springer Singapore, 2019, p.~25–45.

\bibitem{Jurowetzki2021-rm}
{\sc Jurowetzki, R., Hain, D., Mateos-Garcia, J., and Stathoulopoulos, K.}
\newblock The privatization of {AI} research(-ers): Causes and potential consequences -- from university-industry interaction to public research brain-drain?

\bibitem{Klinger2020}
{\sc Klinger, J., Mateos-Garcia, J.~C., and Stathoulopoulos, K.}
\newblock A narrowing of ai research?
\newblock {\em SSRN Electronic Journal\/} (2020).

\bibitem{Kotiranta2020}
{\sc Kotiranta, A., Tahvanainen, A., Kovalainen, A., and Poutanen, S.}
\newblock Forms and varieties of research and industry collaboration across disciplines.
\newblock {\em Heliyon 6}, 3 (Mar. 2020), e03404.

\bibitem{Latour1987-ul}
{\sc Latour, B.}
\newblock {\em Latour: Science in action - how to follow scient ists \& engineers through society (cloth)}.
\newblock Harvard University Press, London, England, July 1987.

\bibitem{latour2013laboratory}
{\sc Latour, B., and Woolgar, S.}
\newblock {\em Laboratory life: The construction of scientific facts}.
\newblock Princeton university press, 2013 [1979].

\bibitem{moore2011science}
{\sc Moore, K., Kleinman, D.~L., Hess, D., and Frickel, S.}
\newblock Science and neoliberal globalization: a political sociological approach.
\newblock {\em Theory and Society 40\/} (2011), 505--532.

\bibitem{Noordegraaf2020}
{\sc Noordegraaf, M.}
\newblock Protective or connective professionalism? how connected professionals can (still) act as autonomous and authoritative experts.
\newblock {\em Journal of Professions and Organization 7}, 2 (June 2020), 205–223.

\bibitem{Papatsiba2013}
{\sc Papatsiba, V.}
\newblock The idea of collaboration in the academy: Its epistemic and social potentials and risks for knowledge generation.
\newblock {\em Policy Futures in Education 11}, 4 (Jan. 2013), 436–448.

\bibitem{Perkmann2021}
{\sc Perkmann, M., Salandra, R., Tartari, V., McKelvey, M., and Hughes, A.}
\newblock Academic engagement: A review of the literature 2011-2019.
\newblock {\em Research Policy 50}, 1 (Jan. 2021), 104114.

\bibitem{Raimbault2022}
{\sc Raimbault, B.}
\newblock Cadrage industriel et production de connaissances. le cas de la biologie synthétique en france.
\newblock {\em Sociologie du travail 64}, 4 (Dec. 2022).

\bibitem{Raimbault2023}
{\sc Raimbault, B.}
\newblock Faire avec l’industrie: Repenser la crédibilité scientifique par la preuve de concept.
\newblock {\em Revue d’anthropologie des connaissances 17}, 2 (May 2023).

\bibitem{Rikap2024}
{\sc Rikap, C.}
\newblock Varieties of corporate innovation systems and their interplay with global and national systems: Amazon, facebook, google and microsoft’s strategies to produce and appropriate artificial intelligence.
\newblock {\em Review of International Political Economy 31}, 6 (June 2024), 1735–1763.

\bibitem{Sarewitz2016PublicationFlood}
{\sc Sarewitz, D.}
\newblock The pressure to publish pushes down quality.
\newblock {\em Nature News 533}, 7602 (2016), 147.

\bibitem{Schot2018}
{\sc Schot, J., and Steinmueller, W.~E.}
\newblock Three frames for innovation policy: R\&d, systems of innovation and transformative change.
\newblock {\em Research Policy 47}, 9 (Nov. 2018), 1554–1567.

\bibitem{Shinn2002Helix}
{\sc Shinn, T.}
\newblock The triple helix and new production of knowledge: Prepackaged thinking on science and technology.
\newblock {\em Social Studies of Science 32}, 4 (Aug. 2002), 599–614.

\bibitem{Shinn2002}
{\sc Shinn, T., and Joerges, B.}
\newblock The transverse science and technology culture: Dynamics and roles of research-technology.
\newblock {\em Social Science Information 41}, 2 (June 2002), 207–251.

\bibitem{Shinn2006}
{\sc Shinn, T., and Lamy, E.}
\newblock Paths of commercial knowledge: Forms and consequences of university–enterprise synergy in scientist-sponsored firms.
\newblock {\em Research Policy 35}, 10 (Dec. 2006), 1465–1476.

\bibitem{Smith2023}
{\sc Smith, R.~D., Sch\"{a}fer, S., and Bernstein, M.~J.}
\newblock Governing beyond the project: Refocusing innovation governance in emerging science and technology funding.
\newblock {\em Social Studies of Science 54}, 3 (Nov. 2023), 377–404.

\bibitem{JMLR:v21:20-091}
{\sc Tavenard, R., Faouzi, J., Vandewiele, G., Divo, F., Androz, G., Holtz, C., Payne, M., Yurchak, R., Ru{\ss}wurm, M., Kolar, K., and Woods, E.}
\newblock Tslearn, a machine learning toolkit for time series data.
\newblock {\em Journal of Machine Learning Research 21}, 118 (2020), 1--6.

\bibitem{Trisovic2022}
{\sc Trisovic, A., Lau, M.~K., Pasquier, T., and Crosas, M.}
\newblock A large-scale study on research code quality and execution.
\newblock {\em Scientific Data 9}, 1 (Feb. 2022).

\bibitem{Uzzi2013}
{\sc Uzzi, B., Mukherjee, S., Stringer, M., and Jones, B.}
\newblock Atypical combinations and scientific impact.
\newblock {\em Science 342}, 6157 (Oct. 2013), 468–472.

\bibitem{Wattanakriengkrai2022}
{\sc Wattanakriengkrai, S., Chinthanet, B., Hata, H., Kula, R.~G., Treude, C., Guo, J., and Matsumoto, K.}
\newblock Github repositories with links to academic papers: Public access, traceability, and evolution.
\newblock {\em Journal of Systems and Software 183\/} (Jan. 2022), 111117.

\bibitem{https://doi.org/10.48550/arxiv.2103.06312}
{\sc Zhang, D., Mishra, S., Brynjolfsson, E., Etchemendy, J., Ganguli, D., Grosz, B., Lyons, T., Manyika, J., Niebles, J.~C., Sellitto, M., Shoham, Y., Clark, J., and Perrault, R.}
\newblock The ai index 2021 annual report, 2021.

\end{thebibliography}

\newpage

\section{Supplementary materials}

\subsection{Supplementary material 1 : Sampling and collection procedures} \label{sup1}

The initial set of papers was found on the ''papers with code platform'' (https://paperswithcode.com/). We decided to use a custom classification given the sparsity of platform categorisation at the time and used a keyword method developed by Gargiulo and colleagues \cite{Gargiulo2023}. Each paper was assigned to a single field of AI according to the frequency of its keywords. We then collected each paper's metadata via the OpenAlex API, using the associated arXiv DOI where possible or the paper title when a perfect match was found.   
We then considered the OpenAlex information in relation to the author's institution (at the time of publication) or extracted the author's displayed affiliation (within the paper pdf) using the Grobid software.
We classified each institution as belonging to either academia, industry, or another type of institution (government, non-profit, etc.) using an open database of university names (https://github.com/Hipo/university-domains-list) and a dump from the ''Crunchbase'' platform (https://www.crunchbase.com/). We performed a ''fuzzy matching'' to assign each institution name to the institution type and manually matched the remaining names. 
Finally, we simplified the classification by considering as academic the studies that included only academics or other researchers affiliated with a non-profit organisation, as private the studies conducted only by researchers affiliated with a company, and as ''mixed'', the studies conducted jointly by the two groups.
Using this logic results in a drastic reduction in data size, considering that we discarded large portions of the dataset at each step if the matching process was inconclusive (see the associated figure).

\begin{figure}[H]
    \centering
    \includegraphics[width=0.75\linewidth]{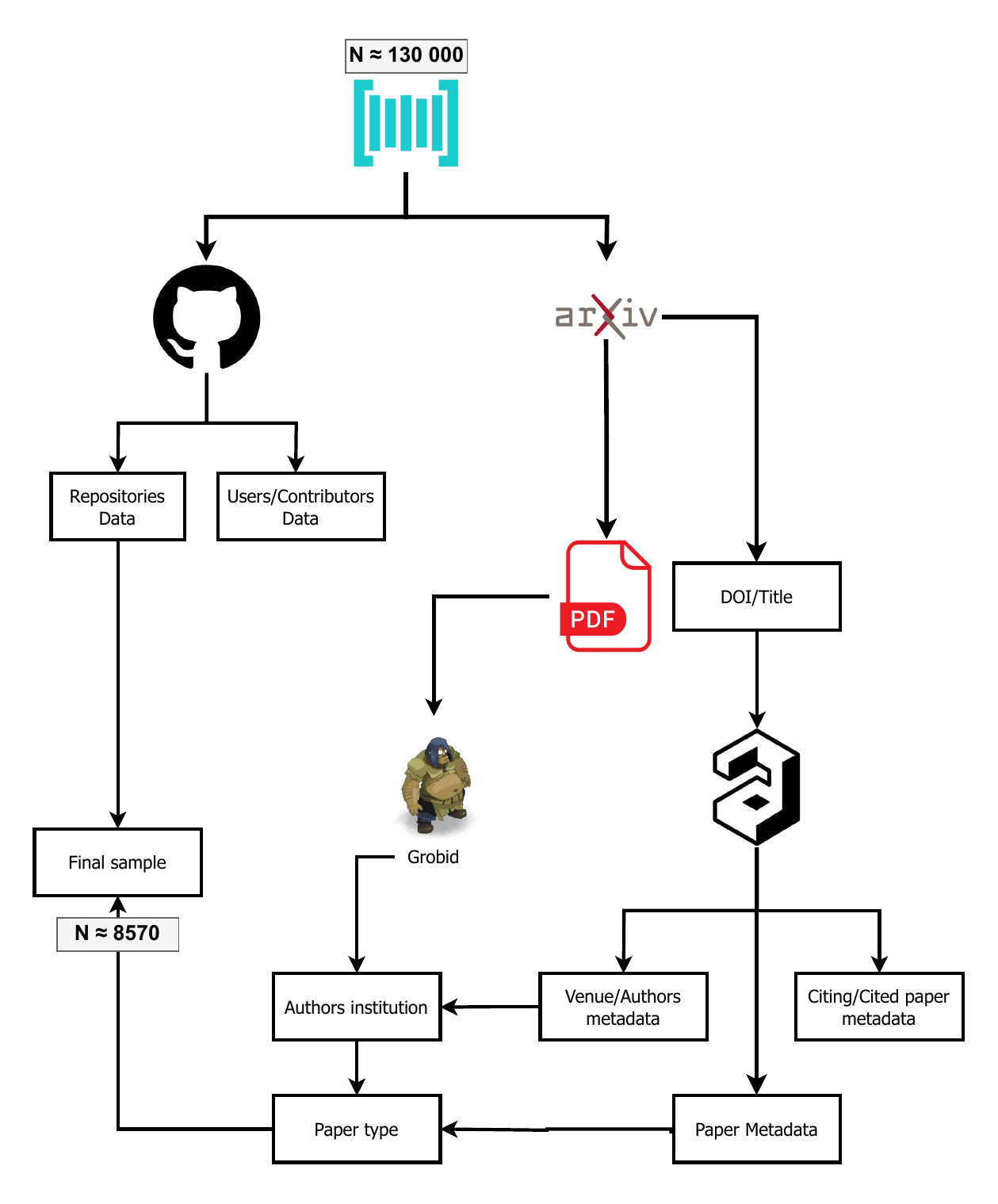}
    \caption{Data collection pipeline. }
    \label{supp_fig_1}
\end{figure}

\subsection{Supplementary material 2 : Programming languages incidence}\label{sup2}

\begin{figure}[H]
    \centering
    \includegraphics[width=0.85\linewidth]{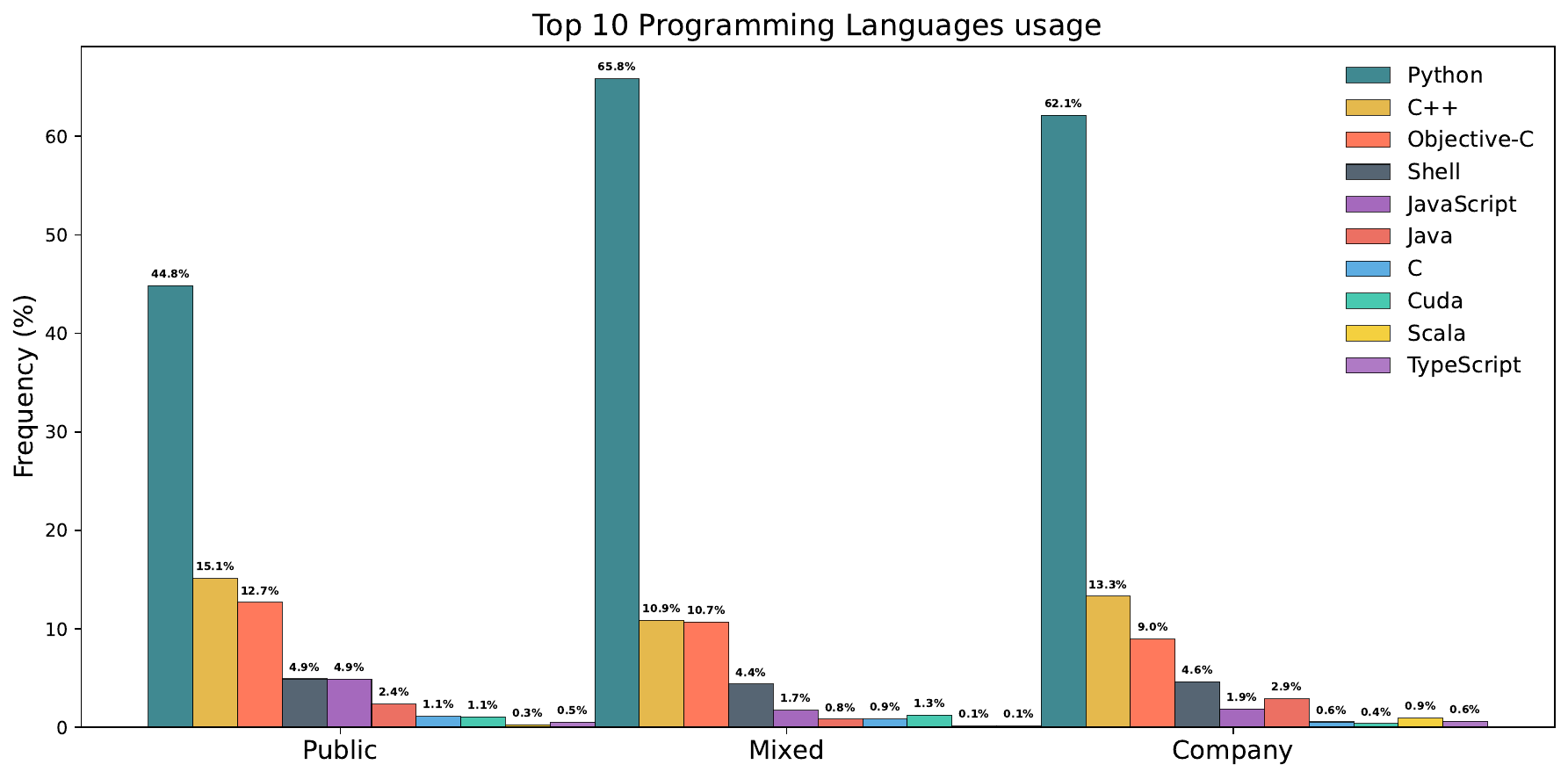}
    \caption{Programming languages frequency across group at the file level.\\ $P_{g}^{\text{all}}(L)= \left\{ l : \frac{\text{lc}_{g}^{\text{all}}(l)}{T_{g}^{\text{all}}} \;\middle|\; l \in L, \; g \in \{a, m, i\} \right\}$. We consider $L$ the set of languages, $l$ a particular element of $L$, $g$ a group of repositories and divide $lc_g(l)$, the total number of the times the programming languages appearing in $g$ by the total count of languages occurrences in $g$.}
    \label{supp_fig_2}
\end{figure}

\begin{figure}[H]
    \centering
    \includegraphics[width=0.85\linewidth]{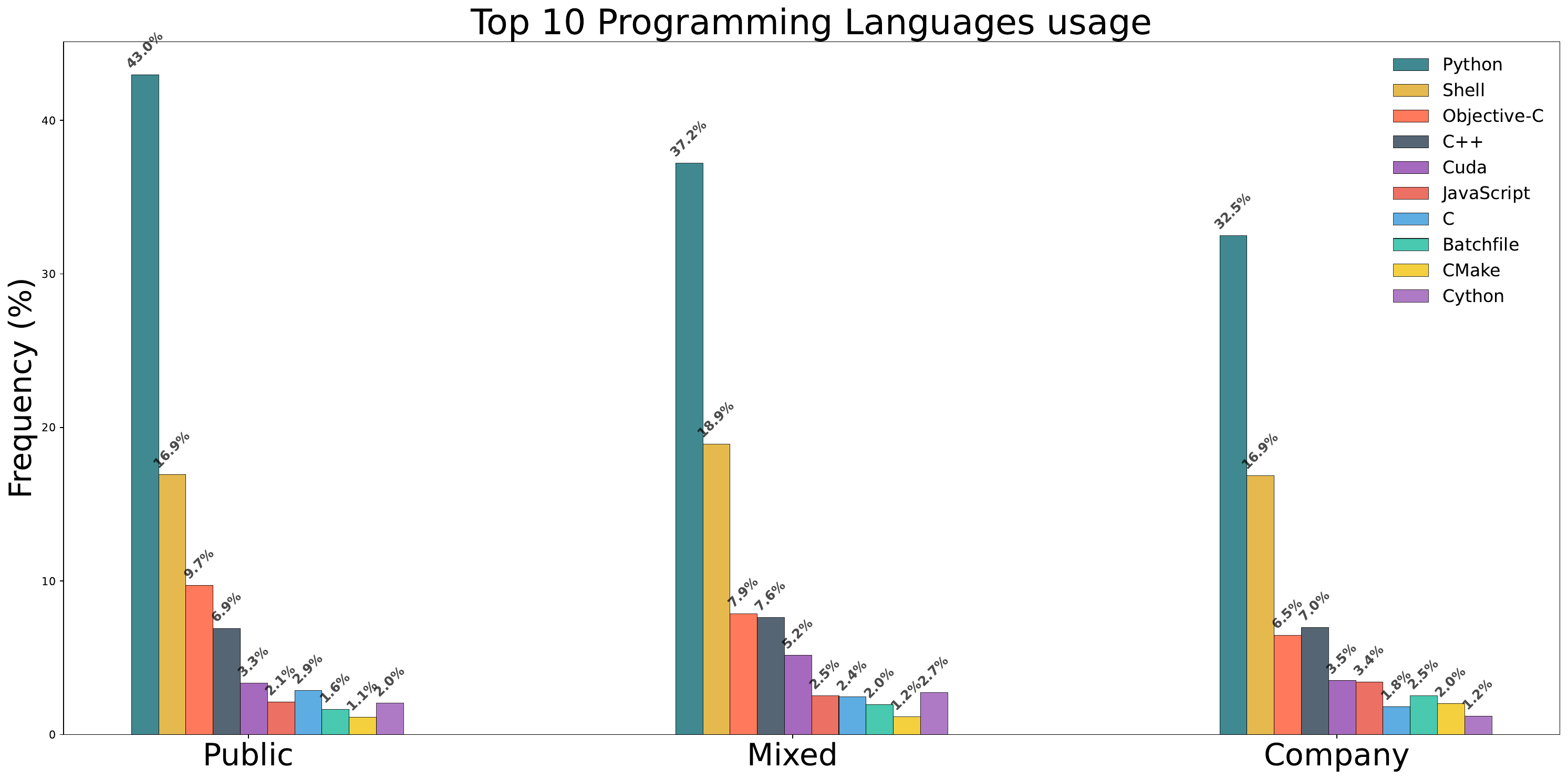}
    \caption{Programming languages frequency across group at the file level.\\ $P_{g}^{\text{set}}(L)=\left\{ l : \frac{\text{lc}_{g}^{\text{set}}(l)}{T_{g}^{\text{set}}} \;\middle|\; l \in L, \; g \in \{a, m, i\} \right\}$. We consider $L$ the set of languages, $l$ a particular element of $L$, $g$ a group of repositories and divide $lc_g(l)$, the set of programming languages appearing in $g$ by the total count of unique languages occurrences in $g$.}
    \label{supp_fig_3}
\end{figure}

The figures show the frequency of programming languages for each group. The first figure shows the value calculated by taking into account the sum of each programming language for each group (thus showing the frequency of appearance of the language within the group). In this case, we can see that Python is more common in mixed and private repositories (62\% and 66\% respectively). The second figure shows the prevalence of languages calculated by taking into account the language set for each repository. Here we see an inverse situation with an over-representation of Python in academic repositories. This situation is explained by the aforementioned higher level of programming language diversity found in private repositories. In other words, private actors almost always use Python, but combine it significantly more often with other programming languages.

\subsection{Supplementary material 3 : Publication venue for academic, mixed and industrial research}\label{sup3}

\begin{figure}[H]
    \centering
    \includegraphics[width=0.85\linewidth]{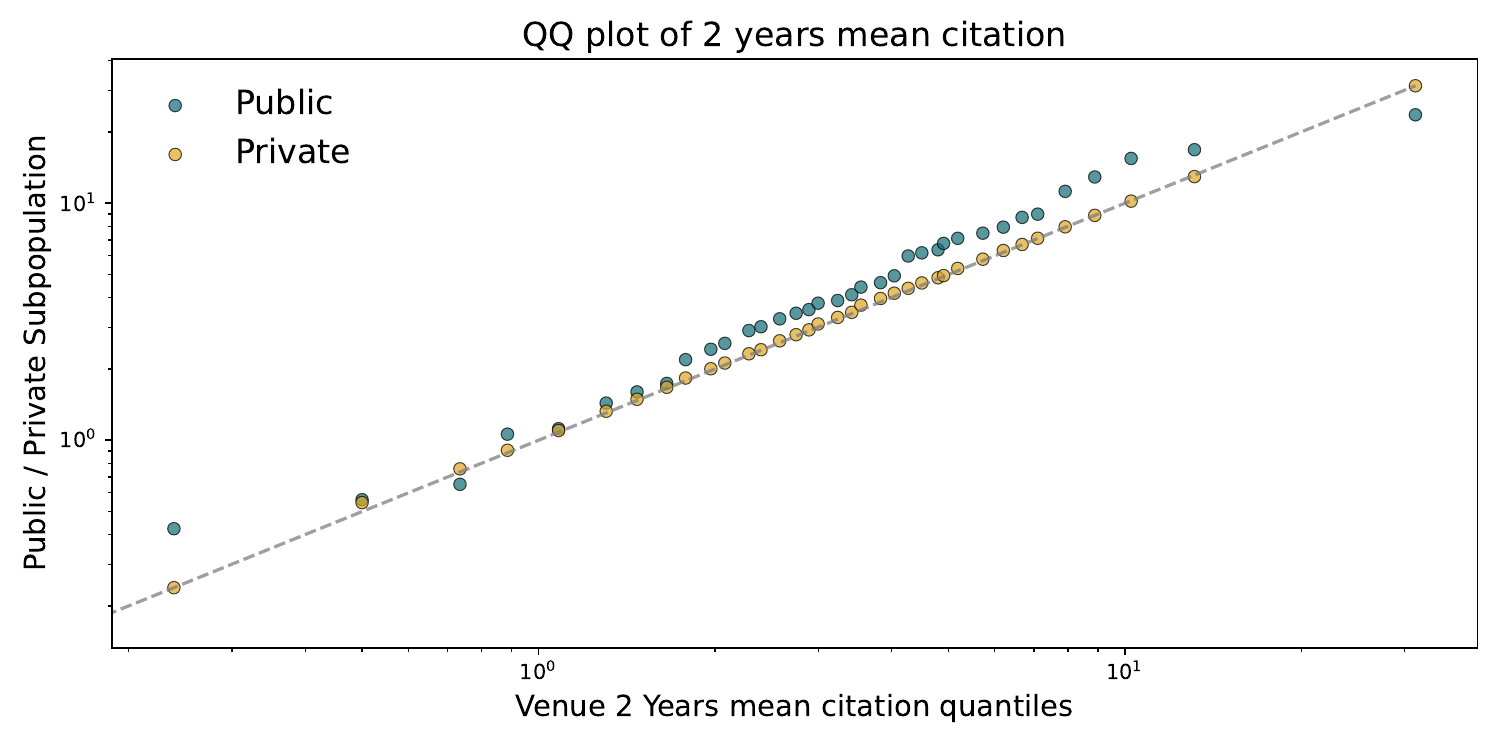}
    \caption{QQ (Quartile to Quartile) plot of the publication venue for Academic and Industrial publication (published articles).}
    \label{supp_fig_4}
\end{figure}

The figures show the quartile distribution of each group venue's mean citation after two years (dots) in comparison to the inter-group quartile distribution. Here we can observe that, even though the distribution between Academia and Industry is very similar, academic actors are over-represented in the lowest quartiles of the distribution and industrials in the upper quartiles. In other words, academics are more likely to go toward low impact journals (with respect to the entire sample) and industrials to the highest impact journals.   

\subsection{Supplementary material 4 : Sampling strategies for Zipf Law}\label{sup4}

To calculate the Zipf law, we only considered repositories whose readme was not empty, and to ensure a fair comparison and check the consistency of the results, we sampled the top 50/100/200 repositories for each group, sorted by stars, number of commits, and number of forks.

\subsection{Supplementary material 5 : Repositories maintenance time}\label{sup5}

\begin{figure}[H]
    \centering
    \includegraphics[width=0.85\linewidth]{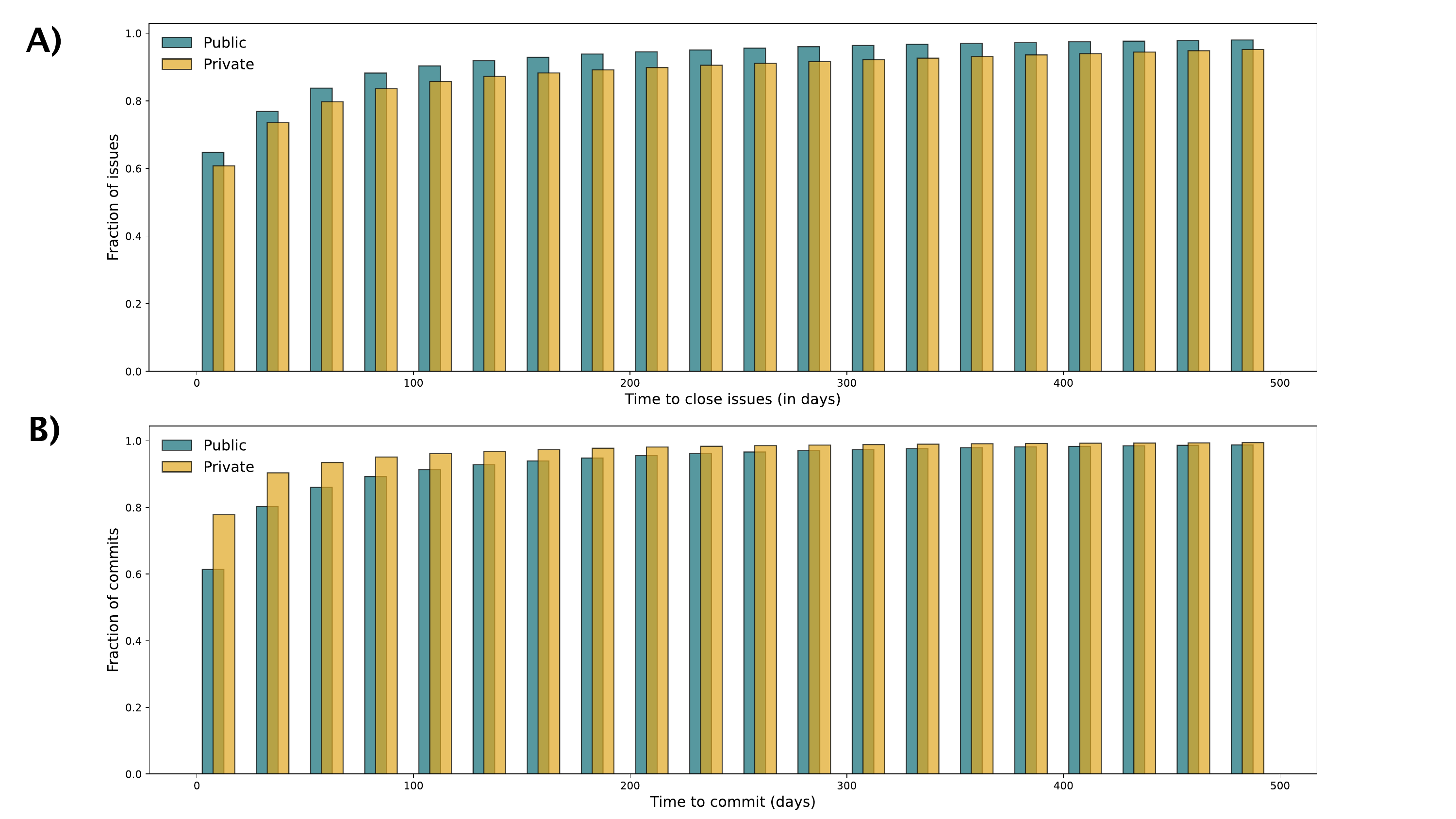}
    \caption{\textbf{Repositories maintenance time} : A) Fraction of issues closed after $n$ days in academic and industrial (at least one industrial authors) repositories B) Time between commits in academic and industrial repositories}
    \label{supp_fig_5}
\end{figure}

The figure shows conflicting trends, with some maintenance tasks being performed more frequently and/or more quickly in industrial repositories and others in academic repositories. We will address these initial findings in future studies, but our initial assessments did not reveal any significant differences between academic and industrial projects.

\subsection{Supplementary material 6 : Cluster selection for times series K-means}\label{sup6}

We used the $\beta \text{cv}$ in order to determine the optimal number of clusters for our clustering. The figure below shows the relation between the number of clusters and the metric. The red line at four for each figure represents the optimal number of clusters. 
This value results from a visual examination of this figure and of the clustering figures. 

\begin{figure}[H]
    \centering
    \includegraphics[width=1.1\linewidth]{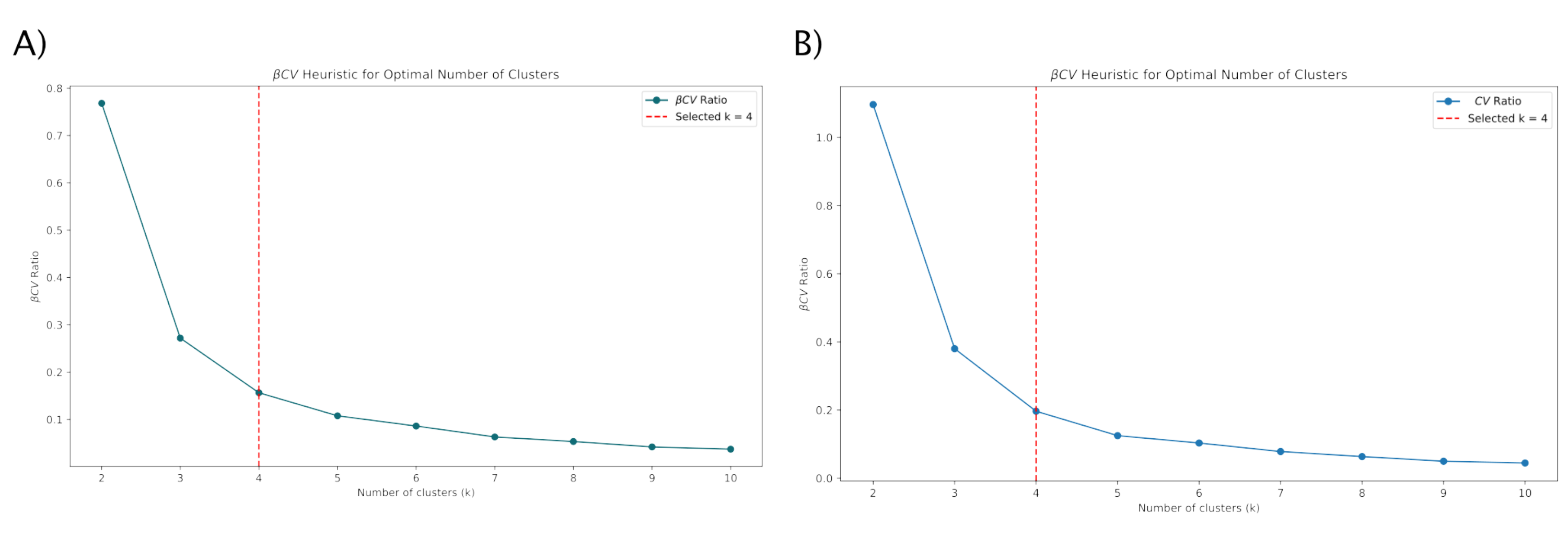}
    \caption{\textbf{Elbow plot for the time series K-Means} : A) Elbow plot for the GitHub stars time series. B) Elbow plot for the citation time series}
    \label{supp_fig_6}
\end{figure}

\end{document}